\documentclass[twocolumn,times]{aastex631}
\expandafter\def\csname editcolor1\endcsname{magenta}
\expandafter\def\csname editcolor2\endcsname{blue}  
\expandafter\def\csname editcolor3\endcsname{violet} 

\usepackage{newtxtext,newtxmath}
\usepackage[T1]{fontenc}
\usepackage{graphicx}	
\usepackage{amsmath}	
\usepackage{float}
\usepackage{bookmark} 
\bookmarksetup{numbered, open,}
\usepackage{enumitem}
\setlist[enumerate]{itemsep=0mm}
\usepackage{hyperref}
\usepackage{subfigure}
\usepackage{xspace}
\usepackage{xcolor} 
\usepackage{soul} 

\usepackage{subfigure} 

\newcommand{\llabel}[1]{\label{#1}}              

\newcommand{\labeq}[2]{ 
\begin{equation} \llabel{#1}
#2
\end{equation}
}

\newcommand{\V}[1]{ {\bf{#1}} }

\newcommand{\pd}[2]{ \frac{ \partial {#1} }{\partial {#2}} }

\newcommand{\D}[2]{ \frac{ d {#1} }{ d {#2} } }

\newcommand{\gcm}{g~cm$^{-3}$\xspace}
\newcommand{\kmps}{\ensuremath{\text{km}~\text{s}^{-1}}\xspace}
\newcommand{\cmps}{\ensuremath{\text{cm}~\text{s}^{-1}}\xspace}
\newcommand{\ctw}{\ensuremath{\text{C}^{12}}\xspace}
\newcommand{\oos}{\ensuremath{\text{O}^{16}}\xspace}
\newcommand{\hef}{\ensuremath{\text{He}^{4}}\xspace}
\newcommand{\netw}{\ensuremath{\text{Ne}^{20}}\xspace}
\newcommand{\mgtf}{\ensuremath{\text{Mg}^{24}}\xspace}
\newcommand{\siteg}{\ensuremath{\text{Si}^{28}}\xspace}
\newcommand{\stt}{\ensuremath{\text{S}^{32}}\xspace}
\newcommand{\arts}{\ensuremath{\text{Ar}^{36}}\xspace}
\newcommand{\cafo}{\ensuremath{\text{Ca}^{40}}\xspace}
\newcommand{\tiff}{\ensuremath{\text{Ti}^{44}}\xspace}
\newcommand{\crfe}{\ensuremath{\text{Cr}^{48}}\xspace}
\newcommand{\feft}{\ensuremath{\text{Fe}^{52}}\xspace}
\newcommand{\nifs}{\ensuremath{\text{Ni}^{56}}\xspace}
\newcommand{\XC}{\ensuremath{\text{X}_\ctw}\xspace}
\newcommand{\XO}{\ensuremath{\text{X}_\oos}\xspace}
\newcommand{\XSi}{\ensuremath{\text{X}_\siteg}\xspace}

\received{today}
\revised{today}
\accepted{today}
\submitjournal{The Astrophysical Journal}

\begin{document}

\title{Three-dimensional Structure of Incomplete Carbon-Oxygen 
       Detonations in Type Ia Supernovae}


\author{A. Khokhlov}
\affil{Department of Astronomy and Astrophysics, the Enrico Fermi Institute, and the Computational Institute, The University of Chicago, Chicago, IL 60637, USA}

\author{I. Dom\'inguez}
\affil{Departamento de F\'isica Te\'orica y del Cosmos,   University of Granada, 18071 Granada, Spain}

\author{A. Y. Chtchelkanova}
\affil{U.S. National Science Foundation, 2415 Eisenhower Ave Alexandria, VA 22314, USA}

\author{P. Hoeflich}
\affil{Department of Physics, Florida State University, Tallahassee, FL 32306, USA }

\author{E.~Baron}
\affil{Planetary Science Institute, 1700 East Fort Lowell Road, Suite 106,
             Tucson, AZ 85719-2395 USA}
\affil{Hamburger Sternwarte, Gojenbergsweg 112, 21029 Hamburg, Germany}

\author{K. Krisciunas}
\affil{George P. and Cynthia Woods Mitchell Institute for Fundamental Physics \& Astronomy, Texas A\&M University, Department of Physics \& Astronomy, 4242 TAMU, College Station, TX 77843, USA}

\author{M. Phillips}
\affil{Las Campanas Observatory, Carnegie Observatories, Casilla 601, La Serena, Chile}

\author{N. Suntzeff}
\affil{George P. and Cynthia Woods Mitchell Institute for Fundamental Physics \& Astronomy, Texas A\&M University, Department of Physics \& Astronomy, 4242 TAMU, College Station, TX 77843, USA}

\author{L. Wang}
\affil{George P. and Cynthia Woods Mitchell Institute for Fundamental Physics \& Astronomy, Texas A\&M University, Department of Physics \& Astronomy, 4242 TAMU, College Station, TX 77843, USA}


\begin{abstract}
Carbon-oxygen (CO) detonation with reactions terminating either after
burning of \ctw in the leading $\ctw + \ctw$ reaction or after
burning of \ctw and \oos to Si-group elements may occur in
the low-density outer layers of exploding white dwarfs and be
responsible for the production of intermediate-mass elements
observed in the outer layers of Type~Ia supernovae.  Basic
one-dimensional properties of CO-detonations have been summarized in
our previous work.  This paper presents the results of two- and
three-dimensional numerical simulations of low-density CO-detonations
and discusses their multidimensional stability, cellular structure,
and propagation through a constant low-density background.
We find three-dimensional CO detonations to be strikingly different
from their one-dimensional and two-dimensional
counterparts.  Three-dimensional detonations are
significantly more robust and capable of propagating without decay
compared to highly unstable and marginal one- and
two-dimensional detonations. The detonation cell size and whether
burning of \ctw in a three-dimensional detonation wave is followed
by the subsequent \oos burning is sensitive to both the background
density and the initial \ctw to \oos mass
ratio.
We also discuss the possible implications for understanding the observed early time bumps in light-curves.

\end{abstract}


\section{Introduction}
\label{Preamble}

Thermonuclear supernovae, so-called Type Ia supernovae (SNe Ia), are
the explosive destruction of CO white dwarfs (WDs) in binary or triple
systems. They produce at least 50\%  of the iron-group elements in the
Universe, are laboratories of modern  and plasma physics
\citep{Hoyle_etal_1960}, and provide a yardstick for modern
cosmology. 

The extraordinary discoveries in cosmology overshadow the  diversity of
SNe Ia that has emerged in recent years, leading to 
three main explosion scenarios \citep{hk96}. The first of these considers a surface He detonation (HeD)  that triggers a detonation of a sub-Chandrasekhar mass (sub-$M_{\rm Ch}$) WD with a low-density CO core \citep{wwt80,n82,hk96}. A second scenario considers a delayed detonation (DD) \citep{khok89} in which the explosion of a  near- $M_{Ch}$-mass WD is triggered by compressional heating at high densities, resulting in the production of electron-capture (EC) elements. Here, the flame propagation starts as a deflagration and transitions to a detonation \citep{Khokhlov1989}.  
In the third leading scenario, two WDs merge or collide in a binary or triplet system, possibly head-on,  \citep{Webbink84}, causing large-scale asymmetries 
and low-density burning. 

Despite all efforts, none of the current models are based on `first principles'. The fact that nuclear physics determines the structure of the WD, the energy release, and the production of elements during the explosion leads to similar spectra and light curves and causes, to first order, a `stellar amnesia' of the progenitor system  despite the expected potential diversity in explosions and progenitor channels. 

Advances in observations during the last decades are starting to overcome the degeneracy in the solutions, but have also exposed a lack of understanding of details and the need for physical effects not considered previously, such as the cellular detonation structure in incomplete carbon burning, which is the  topic of this paper. 
For a collection of reviews by several groups see \cite{2017hsn..book.....A}. 

The outermost layers of the ejecta ($\approx 10^{-5}$ to $10^{-2} M_\odot$) hold a key to understanding the nature of thermonuclear SNe, the progenitor systems, and their environment. They are exposed to the emergent radiative flux  during the first few days after the explosion \citep[see Fig. 3 of][]{Hoeflich_etal_2019}.
 They involve three main components: (1) the surface layers of the exploding WD, 
 (2) all other outer matter bound in the progenitor system, which may include a companion star, and any bound circumstellar matter (CSM)  such as the accretion disc, and the inner parts of several winds (e.g., from the WD, a companion star and an accretion disc), and (3) the unbound CSM consisting of the outer faster and less dense parts of the winds and the interstellar medium. 

Only with recent advances in time-domain astronomy, specific features related to the surface layers have been observed. These include 
early bumps in the light curves and colors \citep{Marion_etal_2016,Hosseinzadeh_etal_2017,Contreras_etal_2018,2019Dimitriadis,2019ApJ...870...13S,2022arXiv220811201B,2022arXiv220707681B,2022MNRAS.512.1317D,2022ApJ...933L..45H,2023ApJ...959..132N,2023ApJ...946....7N,2024ApJ...962...17W}
and the early blue-red-blue evolution
\citep{2023ApJ...959..132N,2022ApJ...933L..45H} which may be
understood in terms of additional energy deposition due to interaction with the outer matter  
or to nuclear burning during the breakout
\citep{2009ApJ...705..483H,2022ApJ...927..180P,2022ApJ...933L..45H},
or as the result of expansion effects
\citep{2022ApJ...932L...2A,Hoeflich_etal_1993}, or, like in
SN~2021aefx, a result of an off-center delayed detonation transition
(DDT) \citep{2023ApJ...959..132N}. Moreover, early-time polarization
spectra of SN~2019np revealed a large asymmetry, but with  fluctuating
polarization spectra hinting at the existence of  small-scale instabilities with
scales of 1000~\kmps that may be attributed to inhomogeneous nuclear
burning products \citep{2023MNRAS.520..560H}. 

The work presented here focuses on the surface layers of the WD. Note that surface burning of He can be expected both in double-detonations of sub-${M_{Ch}}$ mass WDs 
triggered by He-detonation  \citep{2021MNRAS.506.4621B} and in the surface H/He rich layers of near $M_{Ch} $  mass WDs triggered by an outgoing detonation front \citep{2009ApJ...705..483H,Hoeflich_etal_2019}. What is missing in the literature are multi-dimensional simulations corresponding to the properties found close to the WD surface. Three previous studies, done in two-dimensions, were performed in the context of nearly Chandrasekhar mass explosions of SNe Ia, although they focused on larger densities at which the transition from deflagration to detonation is expected to occur, $\approx10^7$~\gcm \citep{Boisseau1996, Gamezo1999, Timmes2000}. These two-dimensional studies  paved the way for our three-dimensional simulations, identifying the relevance of  the transverse waves and cellular structures, and  analyzing the needed resolution and dependence on initial conditions.  Recently, a work by \cite{Iwata2024}, submitted to ApJ, has also presented two-dimensional simulations, in this case for He-detonations in sub-Chandrasekhar SNe Ia.

In Section~\ref{Background}, the overall background is given for the underlying flame physics. In \S~\ref{Formulation}, the problem has been formulated, and the input physics is described. In \S~\ref{Simulations}, the results are presented, and in \S~\ref{Conclusions} the conclusions are summarized and linked to new results in the field of SNe~Ia, and the prospects are discussed. 

A. Khokhlov and I. Dominguez studied instabilities under conditions of the outermost layers by an outgoing detonation wave using full 3D, high-resolution simulations covering many orders of magnitude. Employing Alexei's code ALLA \citep{Khokhlov1998}, the simulations are performed for subdomains between $\approx$ 200 to 3200 km.

This work was performed, and most of the paper was written prior to Alexei Khokhlov's untimely death in 2019. Now, new observations and multidimensional simulations make it highly relevant, and it therefore has been completed by the co-authors.



 \section{Background}
 \label{Background}

It is well known that explosive CO burning in 
SN~Ia proceeds through three distinct nuclear kinetic stages  \citep{1986A&A...158...17T, Gamezo1999}.
(1)~\emph{C-burning:} The \ctw nuclei present in the initial CO
mixture react and burn out in the leading $\ctw + \ctw$ nuclear
reaction forming such products as \oos, \netw, \mgtf,
Na$^{23}$, \siteg, \stt, \hef and free neutrons and protons.
(2)~\emph{O-burning:} The initial and newly formed \oos as well as
other products of the $\ctw + \ctw$ reaction burn to Si-group
elements.  (3)~\emph{Si-burning:} The nuclei of the Si-group burn
into the Fe-peak nuclei, where the matter reaches a state of
nuclear statistical equilibrium (NSE).  For the conditions that occur in SNe~Ia, the
time scales of the above three stages, $t_C$, $t_O$ and $t_{Si}$,
respectively, are vastly different, $t_C << t_O << t_{Si}$, and their
absolute values are a strong function of temperature or,
alternatively, of the background density at which burning takes place.
The cause of the density-dependence of the reaction timescales is the
decrease of the specific heat of degenerate matter with increasing density,
which translates into an increase  of the burning
temperature. 

Densities at which different stages of CO burning take place in a
SN~Ia may be roughly estimated by comparing the nuclear kinetic time
scales with the hydrodynamic timescale of a white dwarf explosion,
$t_h\simeq 1$~s. By equating $t_C$, $t_O$ and $t_{Si}$ to $t_h$ one
finds that Si-burning in SN~Ia occurs at densities $\rho>10^7$~\gcm,
O-burning at densities $10^6<\rho<10^7$~\gcm, and 
C-burning at $10^{4}<\rho<10^6$~\gcm.
Within the framework of current hydrodynamical models 
the above property of CO-burning gives a general explanation of one of the most 
important observational properties of SN~Ia - 
the presence of Fe-peak elements in the inner parts of a
supernova where burning is able to proceed to NSE, and the presence of
intermediate mass elements in the outer layers where burning is
incomplete \citep{Khokhlov1991b}. 

While a general SN~Ia scenario relies on the above
mentioned well established basic thermodynamic and nuclear kinetic
properties of degenerate CO matter and a basic hydrodynamical
timescale of the exploding carbon-oxygen WD, details of 
hydrodynamical models depend on many factors which are still not fully
understood, such as the exact initial exploding configuration, details of the deflagration stage, mechanism(s) of the initiation of a detonation wave and a deflagration-to-detonation transition,
multi-dimensional propagation of detonation waves, etc.  Current
full-star simulations of SN~Ia explosions do not have sufficient numerical resolution to resolve the real structure of a detonation wave, and the level of input physics to model these important processes is not sufficiently well known. 

Detonations occurring at densities below $10^6$ ~\gcm are especially
important for the interpretation of early observations of SNe~Ia before
maximum light, which are accumulating rapidly. At these densities, the
detonation may involve only the first stage of carbon
burning. Hereafter, such detonations are referred to as incomplete or
C-detonations. They are the main focus of this study.  In our
previous paper \citep{dominguez2011}, we dealt with
one-dimensional properties of C-detonations.  This paper presents
the results of two- and three-dimensional numerical simulations of
C-detonations. Our multi-dimensional simulations show that
properties of the same detonation wave are strikingly different when
it propagates in one, two, or three dimensions.  This paper combines
and summarizes the results of our one-, two- and three-dimensional
simulations, explains the role of dimensionality in the propagation of the detonation waves, considers different background densities and initial chemical compositions, and discusses the implications of the results to
SN~Ia modeling.


\section{Formulation of the problem}
\label{Formulation}
\subsection{Equations and input physics}
\label{Equations}
To describe a detonation wave we use time-dependent, compressible,
reactive flow Euler equations of fluid dynamics,
\labeq{Euler-rho}{
       \pd{\rho}{t} + \nabla \cdot \left(\rho \, \V{u} \right) = 0,
}
\labeq{Euler-rhov}{
     \pd{\rho \V{u}}{t}  
        + \nabla \cdot \left(\rho \V{u}\V{u}\right) + \nabla P = 0,
}
\labeq{Euler-E}{
 \pd{E}{t} + \nabla \cdot \left(\V{u} (E + P)\right) =  \rho \dot q,
}
\labeq{Euler-rhoY}{
  \pd{\rho \V{\cal Y}}{t} 
     + \nabla \cdot \left(\rho \, \V{u}\, \V{\cal Y} \right)
      = \rho \V{\cal R},  
}
where $\rho$ is mass density, $\V{u}$ is velocity,
$E=\rho(\epsilon+\frac{u^2}{2})$ is energy density, $\epsilon$ is
internal energy per unit mass, and
$\V{\cal{Y}}=\{{\cal{Y}}_1,...,{\cal{Y}}_N\}$ are mole concentrations
of reactants. $\V{\cal{R}}=\{{\cal{R}}_1,...,{\cal{R}}_N\}$ are the
corresponding net reaction rates which are functions of $\V{\cal{Y}}$,
$\rho$, and temperature $T$.  The nuclear energy generation rate is taken to be
\labeq{Euler-q}{
  \dot q = \V{Q} \cdot \V{{\cal R}} = \sum_{i=1}^N Q_i {\cal R}_i,
}
where $\V{Q} = \{Q_1,..., Q_N\}$ are the binding energies of nuclei.

Nuclear kinetics is described by a network which consists
of $N=13$ nuclei \hef, \ctw, \oos, \netw, \mgtf,
\siteg, \stt, \arts, \cafo, \tiff, \crfe,
\feft, and \nifs. The network takes into account 
reactions between $\alpha$-particles and heavier nuclei $\ctw +
\alpha \leftrightarrow \oos + \gamma$, $\oos + \alpha
\leftrightarrow Ne^{20}$, ..., $Fe^{52} + \alpha \leftrightarrow
Ni^{56}$; binary reactions $\ctw + \ctw$, $\ctw + \oos$, $\oos + \oos$ and the triple-alpha reaction $3 \alpha \leftrightarrow \ctw +
\gamma$.  Effective $(\alpha,\gamma)$ reaction rates are calculated as
sums of contributions from $(\alpha,p)(p,\gamma)$,
$(\alpha,n)(n,\gamma)$, and $(\alpha,\gamma)$ reaction channels
involving protons, neutrons, and photons ($p$, $n$, and $\gamma$,
respectively).  Forward reaction rates and partition functions are
taken from compilations of \citet{Fowler1978, Woosley1978,
Thielemann1993}.  Screening corrections are applied to forward
reactions \citep{YakovlevShalybkov1989}, while reverse reactions are calculated from the principle of detailed balance using ion chemical potentials with liquid-phase Coulomb corrections to account for screening of reverse reactions \citep{YakovlevShalybkov1989}. While not mentioned, the same screening corrections were used in the calculations of \citep{dominguez2011}.  The network captures the multi-stage nature
of explosive CO burning and correctly reproduces the time-scales of
C, O, and Si burning stages.  For densities, $\rho_0 \leq 10^6$~\gcm, burning in SNe~Ia does not proceed to Fe-group elements. In these conditions, the network can be truncated. We verified that a less expensive eight-species $\alpha$-network truncated at \arts provides the same results as a complete $\alpha$-network.  The eight-species network was used in most of the simulations.  The equation of state includes contributions from the ideal Fermi-Dirac electrons and positrons with arbitrary degeneracy and relativism, equilibrium Planck radiation, and Boltzmann nuclei (similar to \cite{Timmes2000b} and originally based in \cite{Nadyozhin1974}).
Some details and a comparison with our previous approximations are discussed in Appendix~\ref{YASS}.

\subsection{Numerical method and computational setup}
\label{NumericalSetup}

Equations \eqref{Euler-rho} - \eqref{Euler-rhoY} are solved using an
adaptive mesh refinement (AMR) reactive-flow fluid dynamic simulation
code ALLA \citep{Khokhlov1998}.  The code uses a directionally split
Godunov-type second-order accurate conservative algorithm for solving
the Euler part of the equations.  Numerical fluxes are evaluated using
a monotone linear reconstruction and a Riemann solver.  The code uses
a dynamic cell-by-cell adaptive mesh refinement.  The mesh can be
characterized by minimum and maximum levels of refinement, $l_{min}$
and $l_{max}$, in all simulations we used four levels of refinement, lmin = lmax-3. At refinement level, $l$, the computational cell size is $\Delta x(l) = L/2^l$, where $L$ is the size of a computational
domain.  The mesh is dynamically refined around shocks, contact
discontinuities, and in regions of large gradients of density,
pressure, and composition of \hef, \ctw, \oos, \netw,
\mgtf and \siteg.  Shocks and discontinuities are always
refined to a maximum level (for details see Section 6, Mesh refinement, in \cite{Khokhlov1998}). The operator-split approach is used for
coupling hydrodynamic and nuclear kinetic algorithms.  A method of  integration of stiff equations of $\alpha$-network nuclear kinetic is
described in Appendix~\ref{YASS}.  The current version of the code, ALLA/HSCD, runs in a hybrid OpenMP/MPI mode on up to $\sim 10^5$ cores (we acknowledge Mr. C. Bacon and the staff of the ALCF at the Argonne National Laboratory for their contribution with the code development and scaling).

\begin{figure}
\centerline{
\includegraphics[scale=0.75]{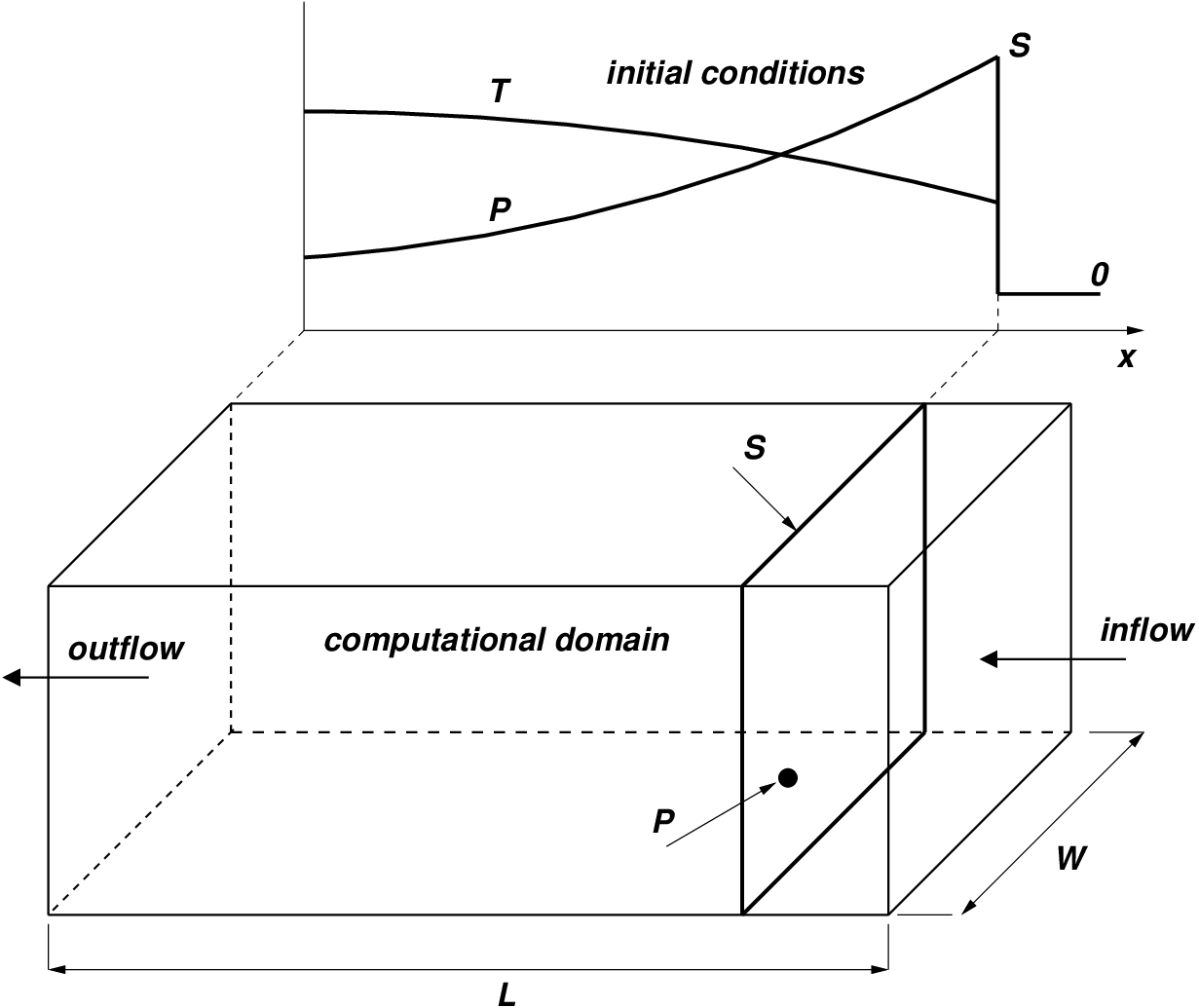}
}
\caption{Three-dimensional computational setup
(Sect.~\ref{NumericalSetup}).  Computational domain is a ``tube'' of length $L$ and cross-section $W\times W$.  The initial steady-state solution (T and P) is mapped onto a mesh with the leading shock, $S$, facing the
right X-boundary. Supersonic inflow with parameters of unburned matter and velocity $-D$ is imposed on the right. Zero-gradient boundary conditions are imposed on the left. Symmetry boundary conditions are
imposed on the Y and Z-boundaries. Spherical perturbation, $P$, is placed close behind the leading shock. Two-dimensional setup is the same as above but with the Z-axis removed.  
\label{Figure-Setup}
}
\end{figure}

Three-dimensional simulations are performed in a
rectangular computational domain (``tube'') with length $L$ and square
cross-section $W^2$ (Fig.~\ref{Figure-Setup}). Steady-state 
Zeldovich-Neumann-D\"oring (ZND) detonation solutions are used as initial conditions.  
The ZND detonation consists of a planar leading shock followed by a one-dimensional reaction zone. 
In a reference frame moving with the detonation velocity, all parameters inside the ZND reaction zone are a function of a distance from the shock, $x$, and do not depend on time. 
ZND solutions are calculated as described in \citep{dominguez2011} and  interpolated onto a computational domain
with the leading shock placed at $x=0.9\,L$ and facing the right
X-boundary.  Simulations are performed in a reference frame moving
with the ZND detonation velocity in order to keep the wave inside the tube.
  A constant supersonic inflow of unburned matter with velocity $-D$,
where $D$ is the detonation velocity, is imposed on the right boundary
and a zero gradient outflow is imposed on the left X-boundary.
Symmetry boundary conditions are used in Y and Z-directions.
Errors of interpolation introduce longitudinal perturbations into initial conditions. A small initial spherical temperature perturbation is added behind the leading shock in order to trigger transverse instabilities.

\subsection{Characteristic scales and numerical resolution}
\label{TimeScales}

A half-reaction length, $x_C$, of a ZND detonation is used in the
paper as a characteristic spatial scale.
It is defined as a distance from the leading
shock to a point where half of the initial \ctw is consumed. The
corresponding temporal scale, the ZND half-reaction time, is 
$t_C=\int_0^{x_C}\,u(x)^{-1}dx$,
where $u(x)$ is the fluid velocity relative to the shock.
The spatial resolution of a simulation can be characterized in
a density-independent way by the number of computational cells of level $l_{max}$ per half-reaction length scale,
\labeq{Delta-x}{
          n_c =  \frac{x_C}{ \Delta x(l_{max})}.
}
The Courant condition on a hydrodynamical time step can then be written as
\labeq{CFL1}{
\Delta t = CFL \cdot \frac{\Delta x(l_{max})}{ \max(a_s + \vert u\vert)} 
         = CFL \cdot \frac{ x_C}{n_c \max(a_s + \vert u\vert)}.
}
CFL is the Courant-Friedrichs-Lewy factor or  Courant factor \citep{Courant1928}, and $a_s$ is the velocity of sound. Since $t_C \simeq x_C/a_s$ it follows that $t_C/\Delta t\simeq n_c$ when $CFL\simeq 1$.
Thus, spatially resolving the reaction zone with the accuracy $\simeq
x_C/n_c$ leads to temporal resolution of the reaction zone with the similar
accuracy $\simeq t_C/n_c$.  In the simulations, we used
$CFL=0.7$ and $n_c = 12.3 - 49$ (Table~\ref{TableConstantRuns3D}).

\section{Results}
\label{Simulations}

\subsection{Input parameters}
\label{InputParameters}

\begin{figure}
\includegraphics[scale=0.9,angle=-90]{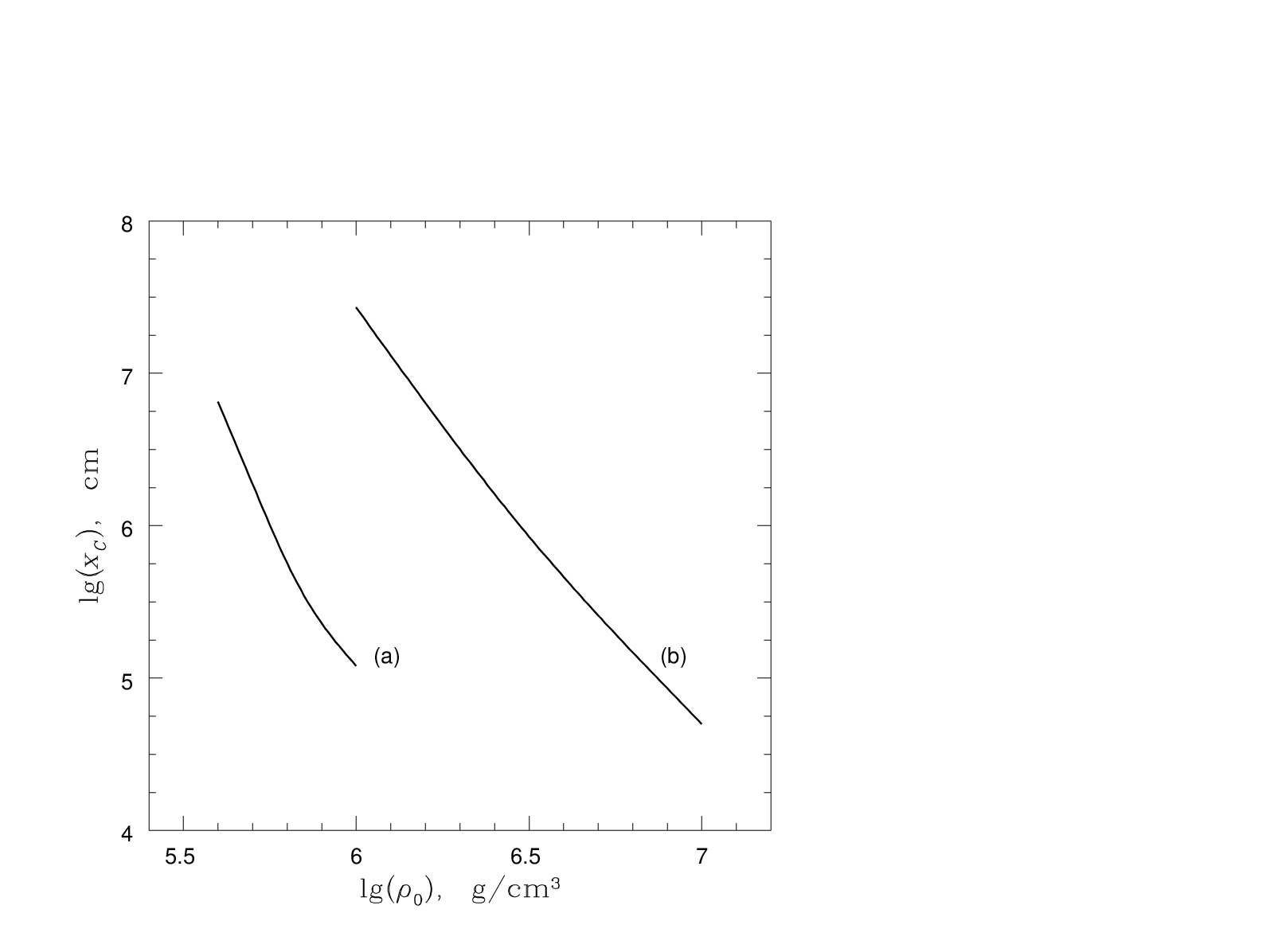}
\caption{
Half-reaction thickness, $x_C$,  of a steady state one-dimensional detonation
as a function of initial density $\rho_0$ for two initial carbon mass fraction, \XC.
(a) $\XC=0.5$ and (b)  $\XC=0.3$.
\label{CJScalesXC}
}
\end{figure}

Simulations were performed for freely propagating Chapman-Jouguet (CJ) detonations. A CJ detonation is characterized by the minimum possible detonation velocity, $D_{CJ}$, the lowest postshock temperature, $T_s$, and the largest $x_C$ and $t_C$ scales as compared to overdriven detonations with velocities $D>D_{CJ}$ (see i.e. \cite{Lee2008} and  \cite{Fickett1979}). Note that here we refer to CJ C-detonation with reduced energy release, not to a CJ complete detonation, in which burning proceeds to NSE (see subsection 3.3 in \cite{dominguez2011}).
We considered two mixture compositions, a $0.5\,\ctw+0.5\,\oos$ 
and a $0.3\,\ctw+0.7\,\oos$ mixture by mass.  
The $0.5\,\ctw+0.5\,\oos$ mixture
is representative of the outer layers of a WD that reaches the
Chandrasekhar mass, accreting H or He from the companion star (single
degenerate scenario). In this case, the C/O ratio is expected to be
$\geq 0.5$ due to radiative shell He burning.  The $0.3\,\ctw+0.7\,\oos$
mixture represents a case when the exploding WD is formed by
merging of two WD (double degenerate scenario). In this case, the
CO accreted matter comes in part from the previous convective He burning core, which is carbon-depleted \citep{Dominguez_etal_2001, Dominguez2003, Bravo_etal_2010}.
Parameters of CJ detonations used as initial
conditions for multi-dimensional simulations are listed in Table.~\ref{TableSteadyState} and Fig.~\ref{CJScalesXC} gives the corresponding half-reaction thickness, $x_C$,  as a function of initial density $\rho_0$ for the two mixture composition considered, \XC=0.5 and 0.3.

\subsection{Three-dimensional detonation wave}
\label{CellularDetonation} 

Table~\ref{TableConstantRuns3D} summarizes three-dimensional
simulations for the $0.5\,\ctw+0.5\,\oos$ composition.  Simulations
are referenced with background density, $\rho_0$, followed by an
alpha-numerical label. For example, $1e6C1$ in
Tab.~\ref{TableConstantRuns3D} indicates a run $C1$ at $\rho_0=10^6$~\gcm.
  For each run, the table gives the length and width of the
computational domain, minimum and maximum levels of refinement, the
numerical resolution, and the location of the initial spherical
perturbation. The obtained results do not depend on the initial perturbations, the boundary conditions, or the size of the computational domain (see also \cite{Boisseau1996} \& \cite{Gamezo1999}).

\begin{figure}
\vskip 1.0cm
\centering
\subfigure[]{
     \includegraphics[scale=0.7,angle=0]{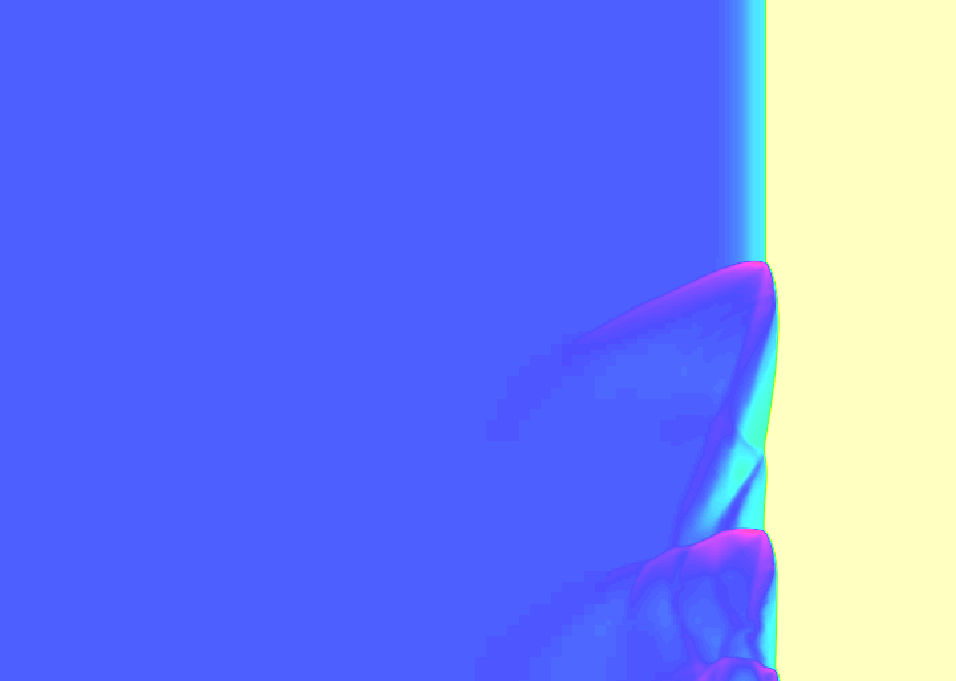}
}
\subfigure[]{
     \includegraphics[scale=0.7,angle=0]{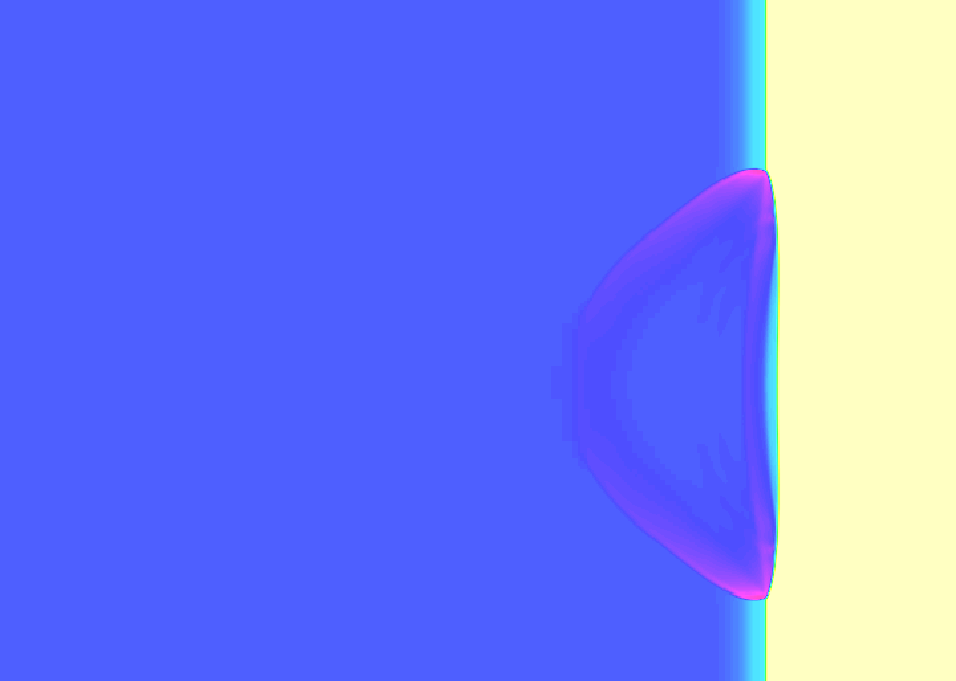}
}
\vskip -16.0cm
\hskip -7.5cm
\includegraphics[scale=0.3,angle=0]{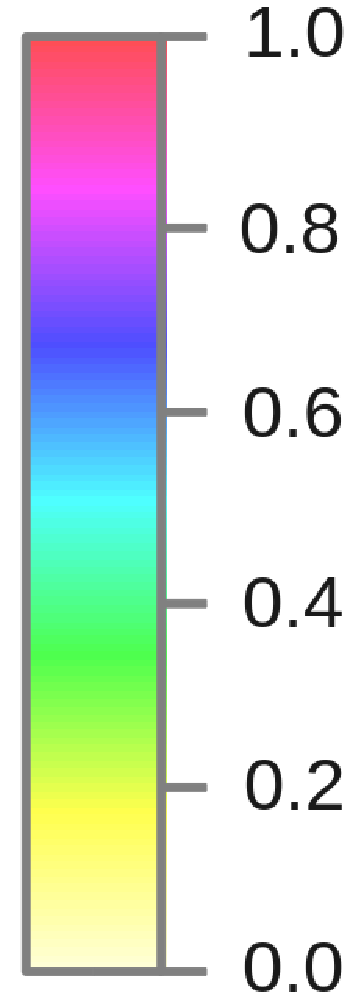}
\vskip 12.0cm
\caption{Initial stage of the development of a cellular Detonation structure 
in the 1e6C2 simulation run.  Temperature is shown in XY (a) and XZ
(b) orthogonal planes passing through the centerline of the
computational domain.  The temperature range is $T=\{0-4\times 10^9\}$K.
Color palette is explained in Appendix~\ref{Visualization}.
\label{Development-1e6C2}
}
\end{figure}

As was explained in Sect~\ref{NumericalSetup} each simulation started with the initially planar one-dimensional ZND detonation wave plus a small spherical temperature perturbation introduced behind the leading shock. A typical development of a multi-dimensional structure of a detonation wave starting from such initial conditions is illustrated in Fig.~\ref{Development-1e6C2} for case $1e6C2$. At the beginning of the simulation, the planar wave starts to decay due to a strong
one-dimensional longitudinal instability.  However, the perturbation
introduced behind the leading shock generates a spherical shock wave
that propagates towards the leading shock and side walls. Reflections
off the walls create additional secondary shocks, which interact with each other and with the leading shock.  The interactions rapidly
trigger a train of transverse reaction waves traveling in orthogonal
directions behind the leading shock.  

Fig.~\ref{Development-1e6C2}a
shows the reflection of a first transverse wave from the bottom wall and
two additional transverse waves traveling toward the upper wall.  An
unaffected part of the planar detonation wave close to the upper wall
is still decaying.  In the orthogonal cross-section
(Fig.~\ref{Development-1e6C2}b) the two transverse waves created by
the interaction of the spherical shock with the leading shock have not
yet reached the boundaries.  The transverse reaction waves accelerate
burning behind the leading shock and prevent the decay of the
detonation wave. Instead of decaying, the detonation wave settles into a
quasi-steady cellular regime of propagation. 

\begin{figure}
\includegraphics[scale=0.7,angle=0]{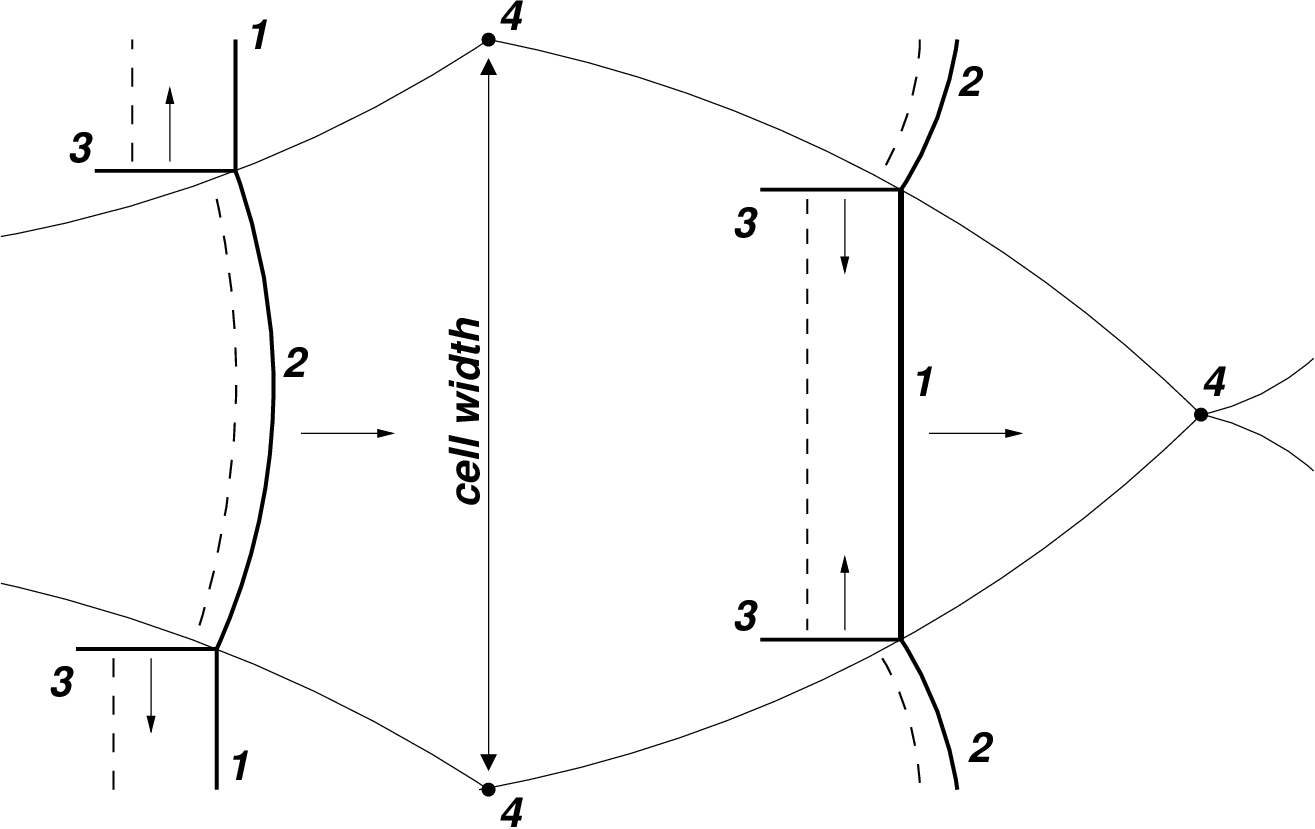}
\caption{Cellular Schematics. Rhomboidal shape of a detonation cell: (1) incident waves, (2) Mach stems, (3) transversal waves and (4) collision points.
\label{CellularSchematics}
}
\end{figure}

The cellular detonation regime is well known \citep{Lee1984,Shepherd2009}
and is schematically illustrated in two dimensions in
Fig.~\ref{CellularSchematics}. The leading shock of a cellular
detonation is not one-dimensional and consists of sections of varying
strength, which are joined at triple points with transverse waves
moving behind the leading shock. The matter that passes through the
leading shock burns in narrow reaction zones behind the strong
sections of the shock and accumulates in thick layers behind the weak
sections of the shock. This matter is subsequently burned in transverse
waves periodically passing through the layers.  Collisions of
transverse waves moving in opposite directions create new
strong sections of the leading shock, whereas previous strong sections
weaken with time. The picture continuously evolves with time and
periodically repeats itself.  Trajectories of triple points form
rhomboidal shapes called detonation cells. In narrow tubes, the
detonation wave may be affected by the walls.  In wide tubes, like those used in these simulations, and in
unconfined detonations, the cell size, $l_c$, depends on the intrinsic kinetic and thermodynamic properties of matter.

The above description is highly
idealized.  Real detonations are three-di\-men\-si\-onal.  Strong and weak
sections of the leading shock are surfaces, and triple points are
triple lines. A regular cellular structure depicted in
Fig.~\ref{CellularSchematics} is only typical of weakly unstable
detonations.  With increasing instability, the cellular structure
becomes irregular and chaotic. In addition, interactions of
secondary waves with the leading shock and with each other generate
shear layers, vortical motions, and turbulence. The relevance of vorticity was first discussed in \cite{Timmes2000} for larger densities.

\begin{figure}
\vskip 1cm
\centering
\subfigure[]{
     \includegraphics[scale=0.7,angle=0]{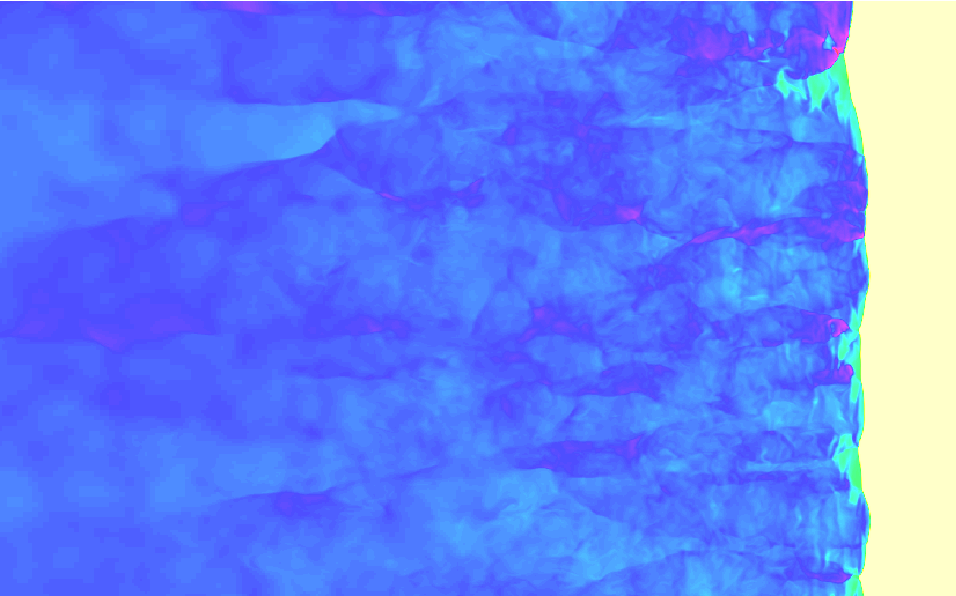}
}
\subfigure[]{
     \includegraphics[scale=0.7,angle=0]{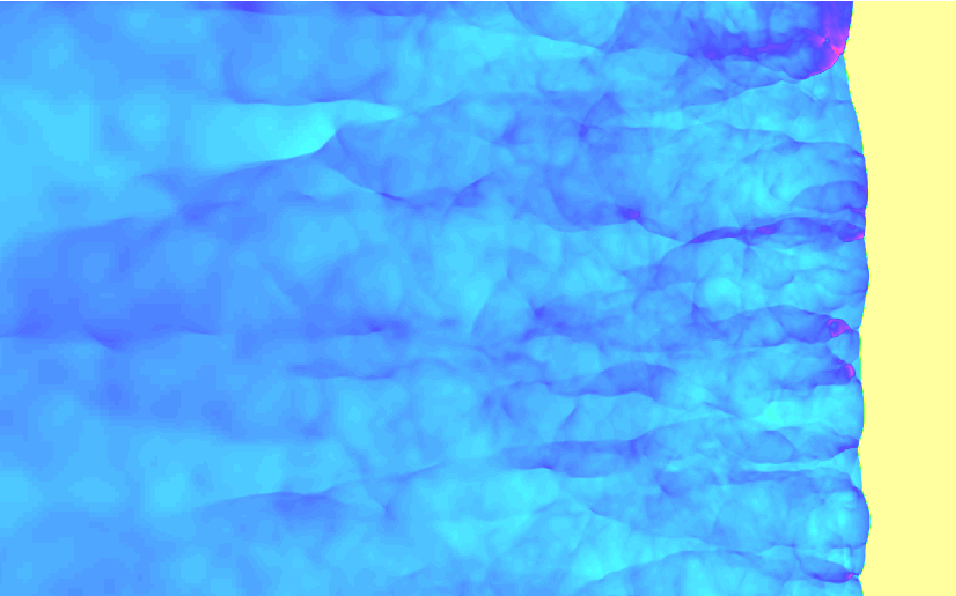}
}
\vskip -15.0 cm
\hskip -7.5cm
\includegraphics[scale=0.3,angle=0]{Scale.eps}
\vskip 11 cm
\caption{(a) $T$ and (b) $\lg P$ for a three-dimensional 
cellular detonation at $\rho_0=10^6$~\gcm (run 1e6C2)
at $t=50 t_c$ in the XZ-plane passing through the centerline of the computational domain. 
The temperature and pressure range shown in the figure is 
$T=\{0-4\times 10^9\}$K and $\lg P=\{22-24\}$~erg~cm$^{-3}$, respectively.
Color palette is explained in Appendix~\ref{Visualization}.
\label{TP-1e6C2}
}
\end{figure}

\begin{figure}
\vskip 1cm
\centering
\subfigure[]{
     \includegraphics[scale=0.7,angle=0]{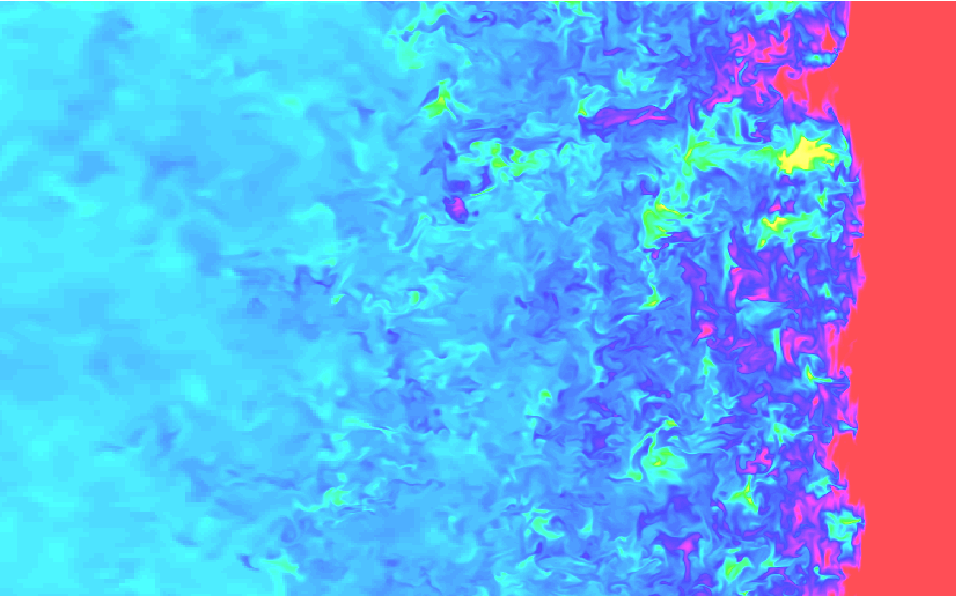}
}
\subfigure[]{
     \includegraphics[scale=0.7,angle=0]{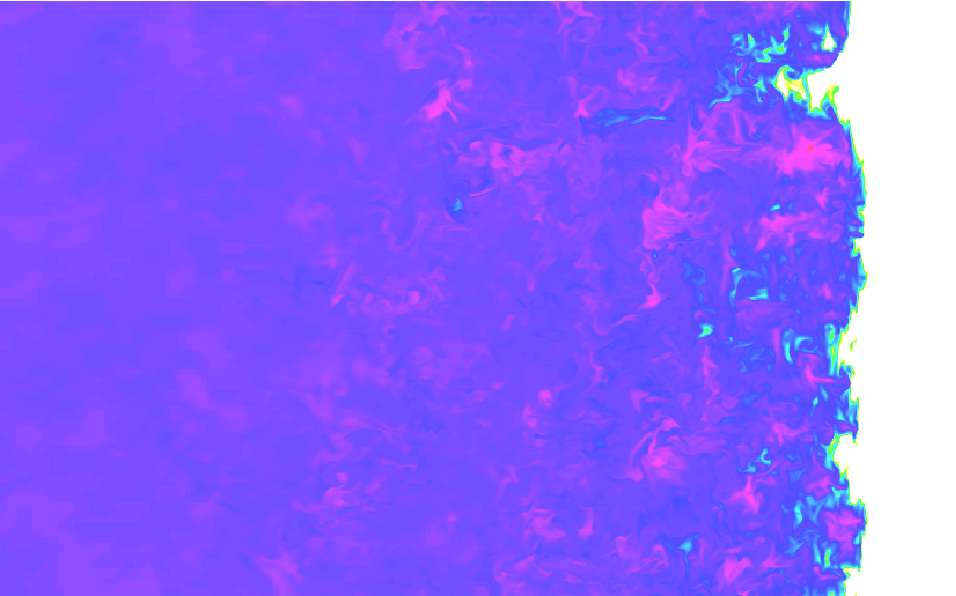}
}
\vskip -15.0 cm
\hskip -7.5cm
\includegraphics[scale=0.30,angle=0]{Scale.eps}
\vskip 10 cm
\caption{Mass fractions (a) $\XC$  and (b) \XSi for a three-dimensional 
cellular detonation at $\rho_0=10^6$~\gcm (run 1e6C2)
at $t=50 t_C$ in the XZ-plane passing through the centerline of the computational domain. 
$\XC=\{0-0.5\}$; $\XSi=\{0-0.6\}$.
Color palette is explained in Appendix~\ref{Visualization}.
\label{CSi-1e6C2}
}
\end{figure}

\begin{figure}
\vskip 1cm
\centering
\subfigure[]{
     \includegraphics[scale=0.7,angle=0]{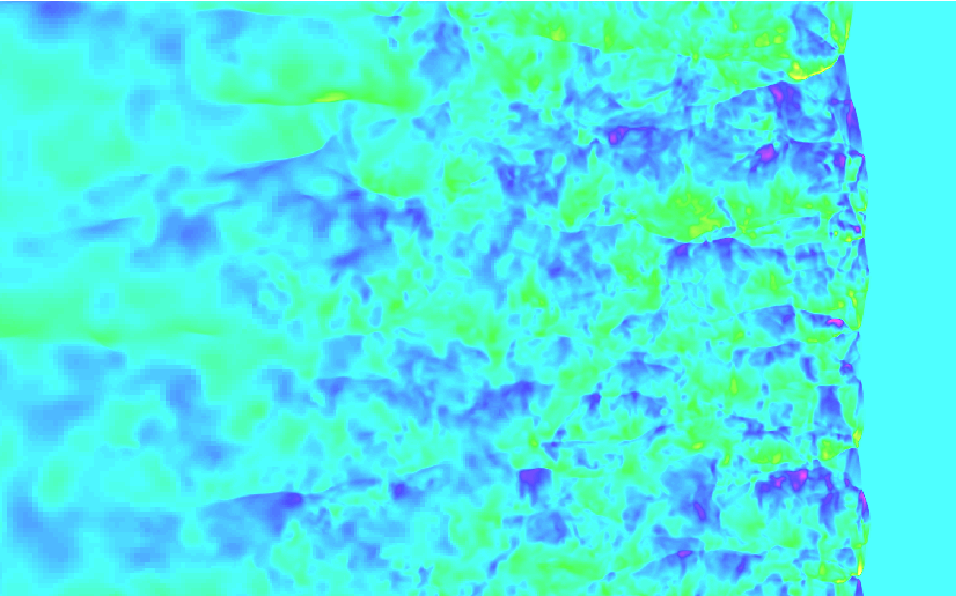}
}
\subfigure[]{
     \includegraphics[scale=0.7,angle=0]{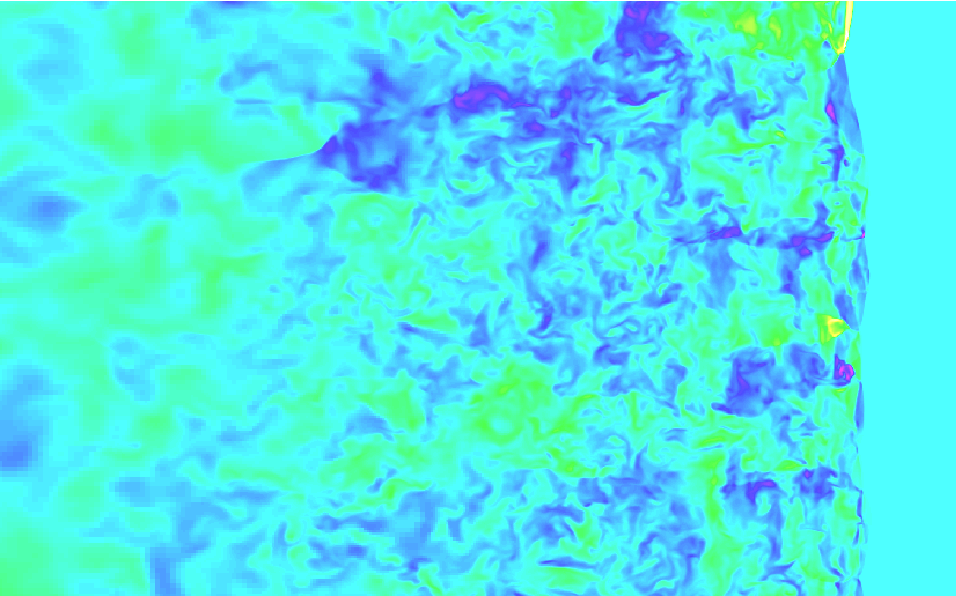}
}
\vskip -15.0 cm
\hskip -7.5cm
\includegraphics[scale=0.30,angle=0]{Scale.eps}
\vskip 10 cm
\caption{Y and Z-components, (a) $u_y$ and (b) $u_z$, of a fluid velocity for a three-dimensional cellular detonation at $\rho_0=10^6$~\gcm (run 1e6C2)
at $t=50t_C$ in the XZ-plane passing through the centerline of the computational domain. Velocity range is $u_{y,z}=\pm\,6\times\,10^8$ \cmps.
Color palette is explained in Appendix~\ref{Visualization}.
\label{VyVz-1e6C2}
}
\end{figure}

The onset of a cellular regime in all simulations occurred approximately after $\simeq 2-3$ sound crossing times of the tube width.  After that,
the detonation propagates in a fully developed cellular regime with
the detonation velocity very close to its Chapman-Jouguet value,
$D\simeq D_{CJ}$.  The simulations show an irregular cellular
propagation of a detonation wave in all three-dimensional runs. As an
example, Figs.~\ref{TP-1e6C2}--\ref{VyVz-1e6C2}
show the cellular structure of the detonation wave in run $1e6C2$ at
$t=50t_c$.  At this time, the detonation is in a fully developed
cellular regime and initial conditions are virtually forgotten.

Strong and weak parts of the leading shock are clearly visible in the temperature plot Fig.~\ref{TP-1e6C2}a.  The temperature is as low as
$\simeq 1.5\times 10^9$~K behind weak parts of the shock and reaches
$\simeq 4.0\times 10^9$~K behind the strong parts.  The steady-state
post-shock temperature is $T_s=2.27\times 10^9$~K, so that temperature
fluctuations behind the leading shock are of the order $\simeq 2$.
Transverse waves are clearly visible on the pressure plot in 
Fig.~\ref{TP-1e6C2}b. The waves are stronger inside the layers of
shocked, unburned matter where they are essentially transverse
detonation waves.  The waves continue as transverse shocks in burned
matter away from the leading shock. Pressure behind the transverse
waves in burned matter varies by a factor of $\simeq 2$. 

Fluctuations
of thermodynamical parameters lead to variations of chemical
composition of the burned matter.  Fig.~\ref{CSi-1e6C2} shows order
$\simeq 100$\% fluctuations of \ctw and \siteg mass fractions
close behind the leading shock.  Fig.~\ref{VyVz-1e6C2} shows
transversal components of fluid velocity.  Fluctuations of the
velocity components on average, are of the order
$\delta\simeq\pm3\times 10^8$~\cmps or $\simeq\pm0.5a_s$.  The largest
fluctuations reach $\simeq a_s$. Close inspection of the results shows
that fluctuations are partially associated with the instantaneous
location of transverse waves and represent the post-shock fluid
velocity. Another component of the velocity fluctuations is associated
with shear layers, interactions of shock waves with density
inhomogeneities, and shock-shock interactions. The two components of
the velocity field are of the same order.

\begin{figure}
\centering
\vskip -5.0cm	
    \includegraphics[scale=0.6,angle=-90]{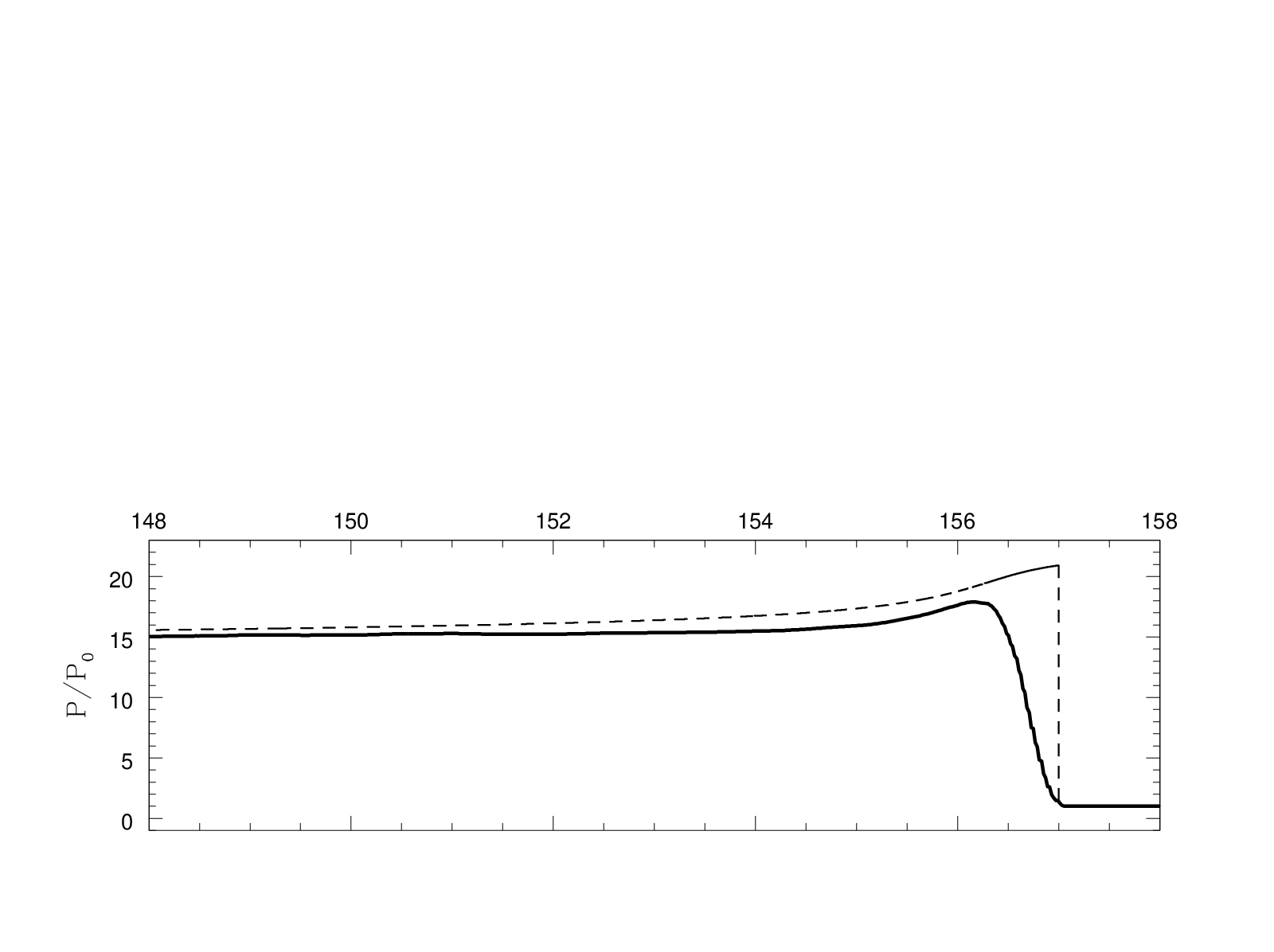}
\vskip -8.3cm
    \includegraphics[scale=0.6,angle=-90]{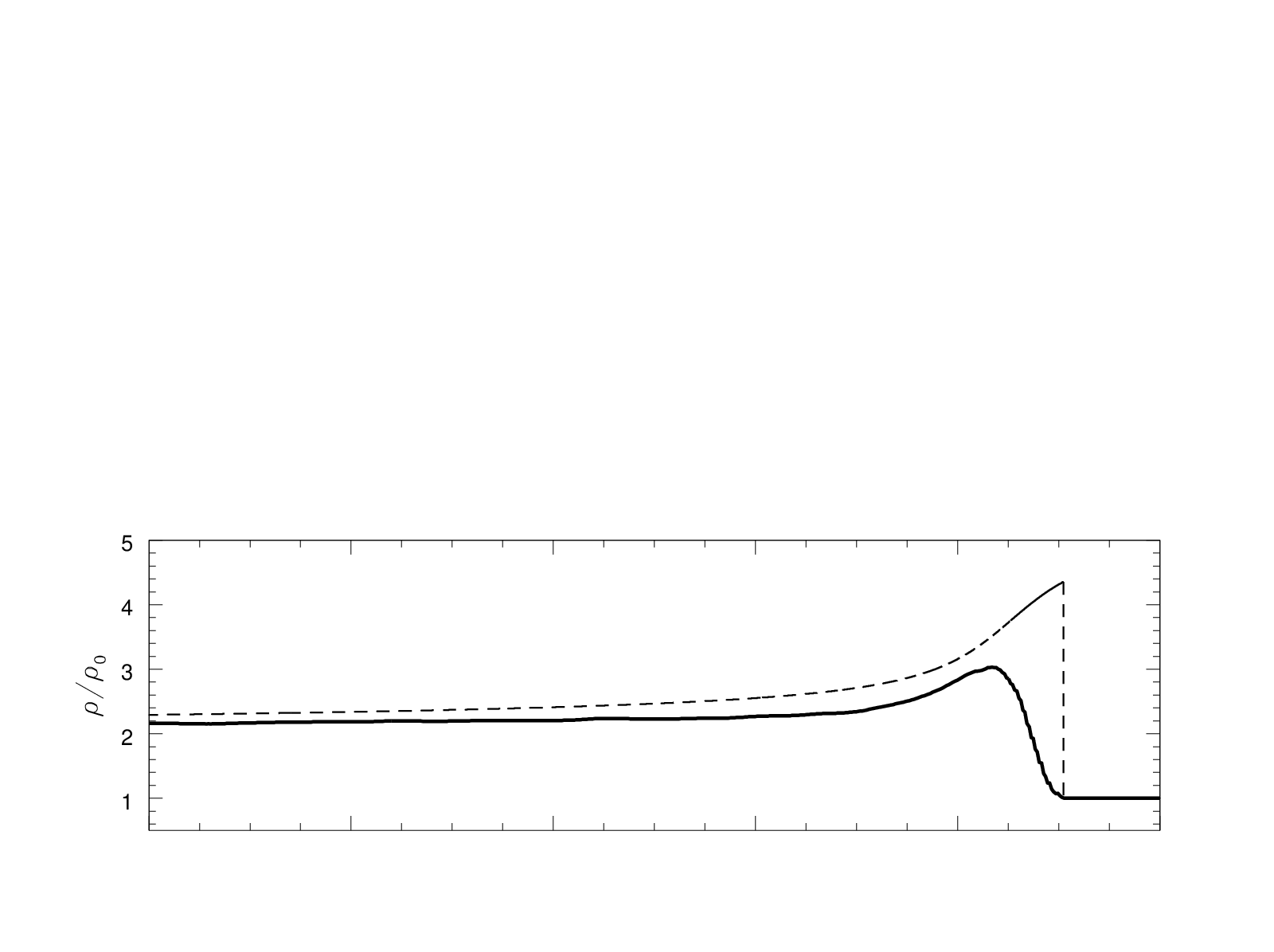}
\vskip -8.3cm
    \includegraphics[scale=0.6,angle=-90]{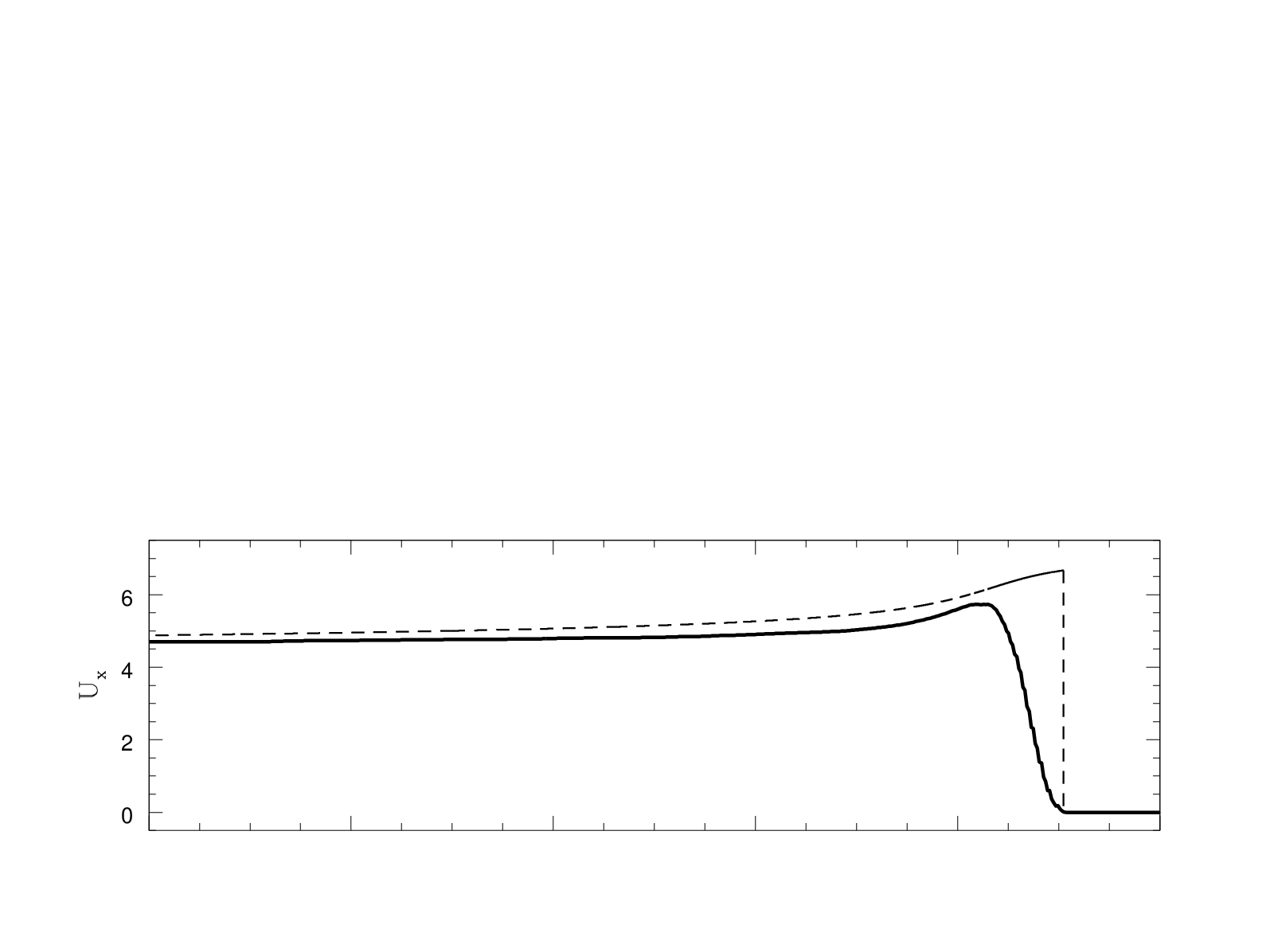}
\vskip -8.3cm
    \includegraphics[scale=0.6,angle=-90]{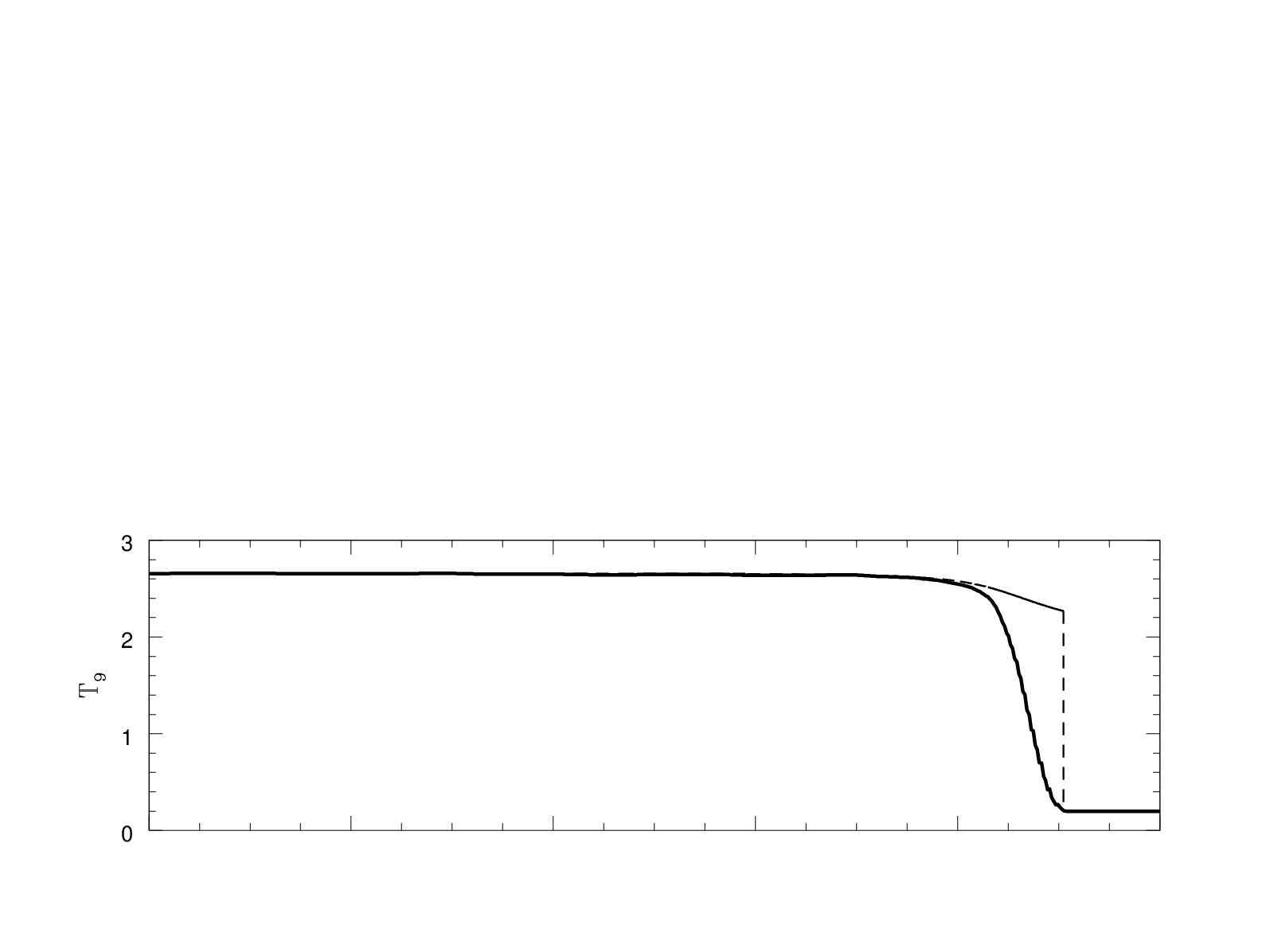}
\vskip -7.2cm
    \includegraphics[scale=0.6,angle=-90]{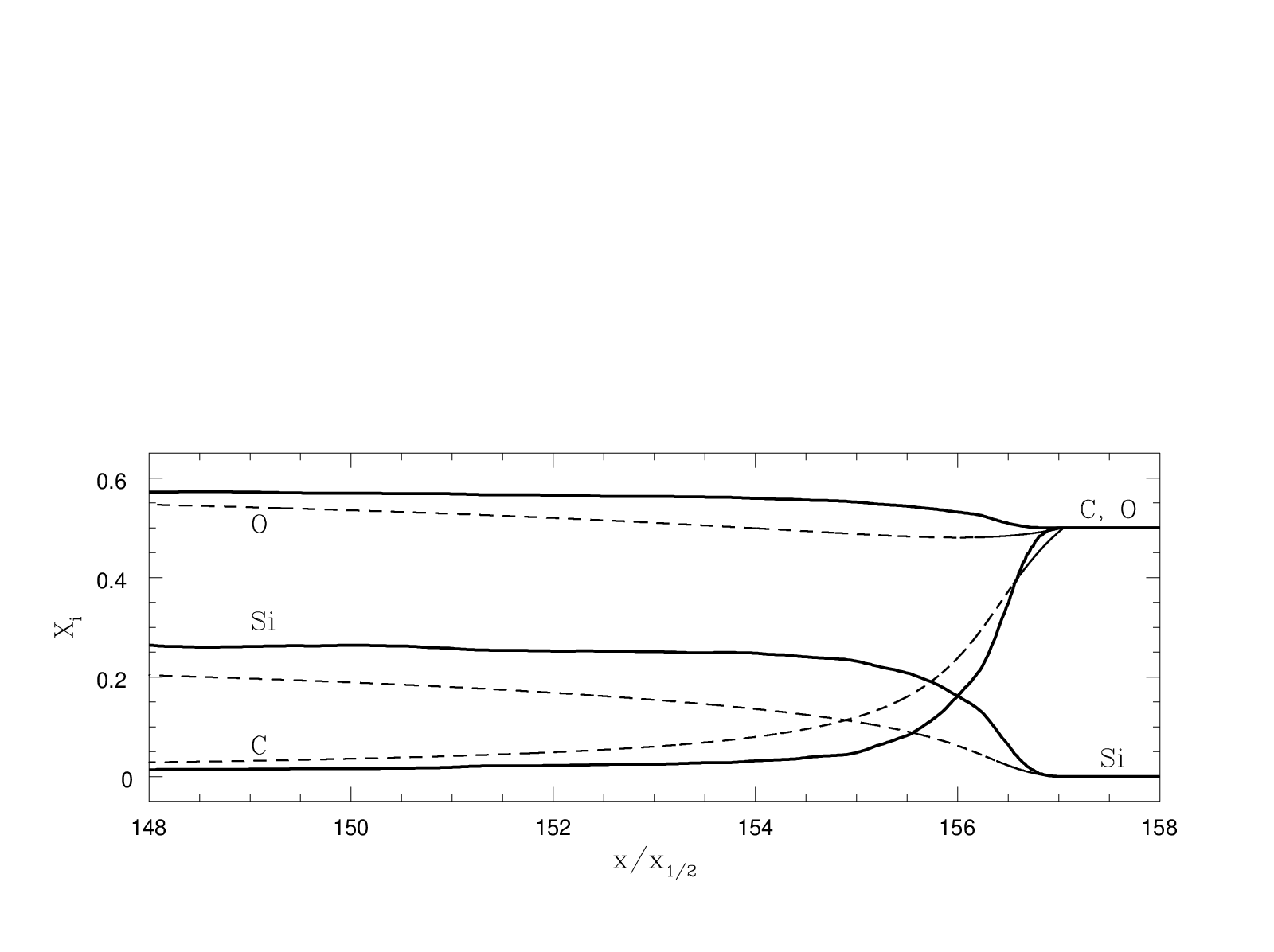}
\vskip -0.7cm
\caption{Averaged one-dimensional structure of a three-dimensional cellular CJ 
detonation at $\rho_0=10^6$~\gcm (run 1e6C2) at $t=50t_C$.
Top to bottom: (1) $P$, (2) $\rho$, (3) $u_x/10^8$ \cmps, (4) $T/10^9$K, 
(5) \XC, \XO, and \XSi  as a function of X-coordinate.
 Dashed lines show the corresponding steady-state solution (initial conditions).
\label{Figure-Average-1e6C2}
}
\end{figure}

Fig.~\ref{Figure-Average-1e6C2} compares a YZ-averaged
three-dimensional structure of run 1e6C2 at $t=50t_C$ with the
corresponding steady-state ZND detonation. The figure shows that the
averaged three-dimensional structure closely resembles the
steady-state one-dimensional reaction zone.  The only difference is
that \ctw is incinerated in a three-dimensional detonation
slightly more rapidly, and slightly more \siteg is produced.  While
slowly decreasing in amplitude, the variations of pressure,
temperature, and velocity around the average values persist in burned
matter long after the matter passes through the leading shock.  The
velocity structures and transverse waves are present behind the
leading shock at distances $\simeq 20-30 x_c$ and more.  The same is
true for chemical non-uniformities of Si, $Ne$, $Mg$ and other
products of C-burning, although the compositional non-uniformities
are typically less than $50$\% of the average and diminish with the
distance from the shock. Compositional non-uniformities of \ctw
rapidly burn out and are generally confined within a short distance
$\simeq 5 x_c$ from the leading shock.  Virtually no carbon remains at
$x > 5 x_c$.  Overall, the three-dimensional detonation appears very
robust.  Similar behavior was found in all other
three-dimensional simulations.

\subsection{Detonation cells}
\label{CellSize}

\begin{figure}
\subfigure[]{
    \includegraphics[scale=0.56,angle=0]{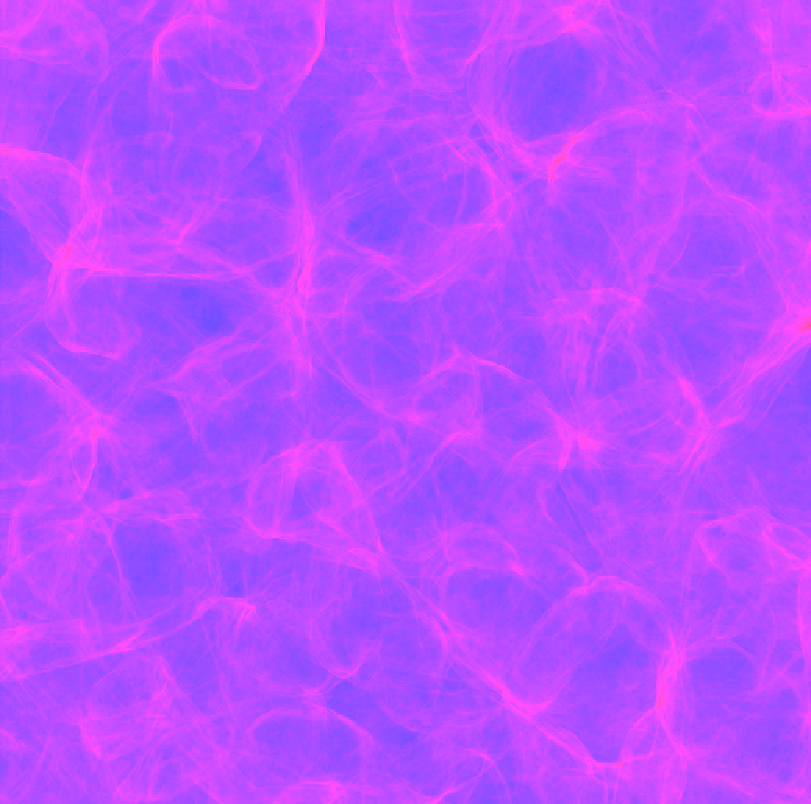}
}
\subfigure[]{
    \includegraphics[scale=0.56,angle=0]{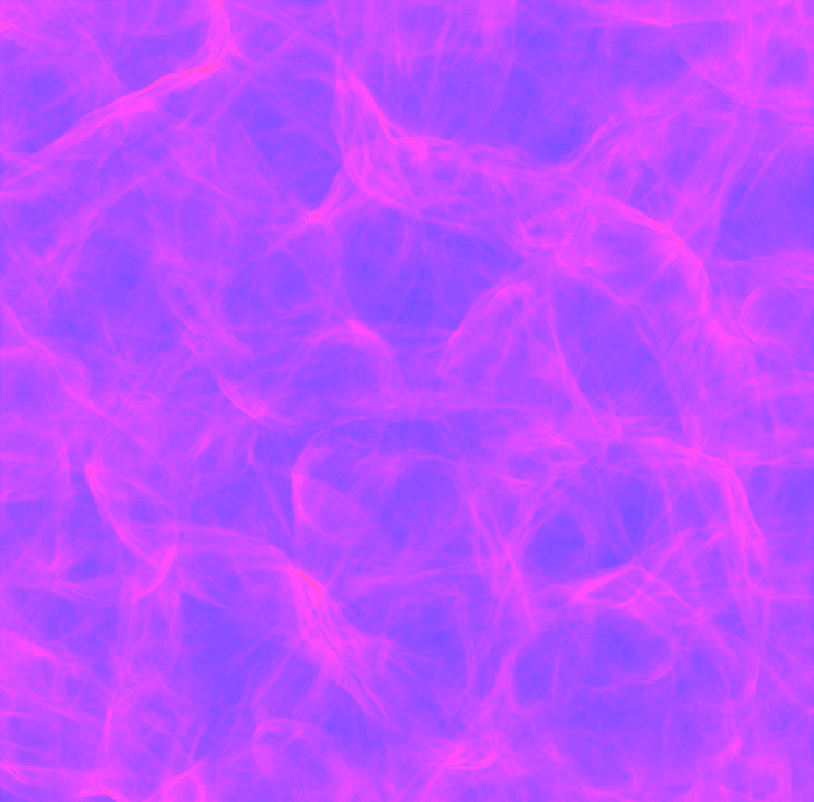}
}
\subfigure[]{
    \includegraphics[scale=0.56,angle=0]{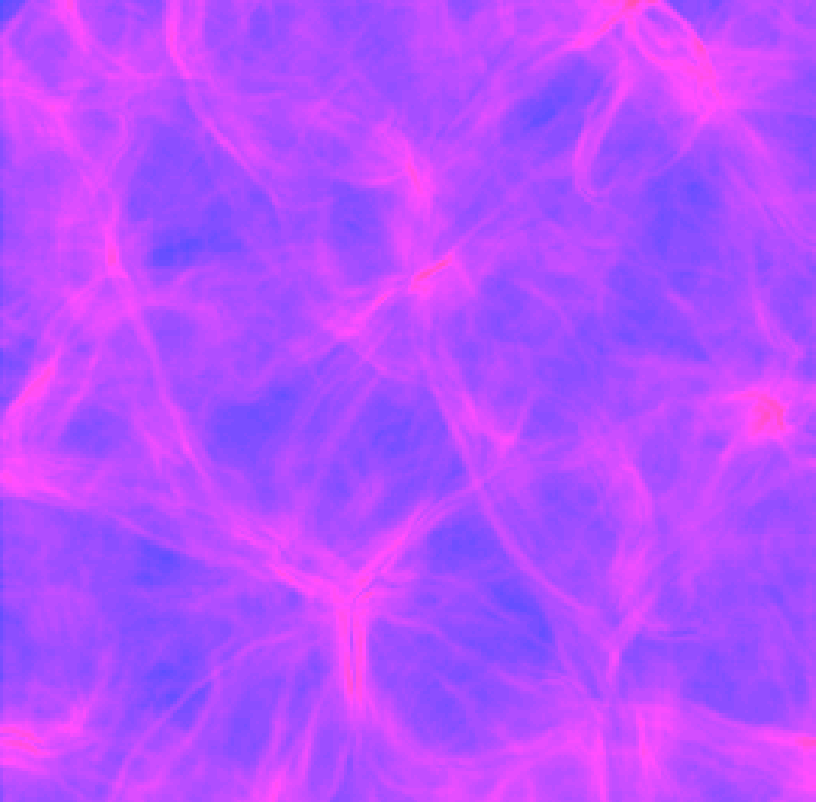}
}
\hskip 2cm
\subfigure[]{
    \includegraphics[scale=0.56,angle=0]{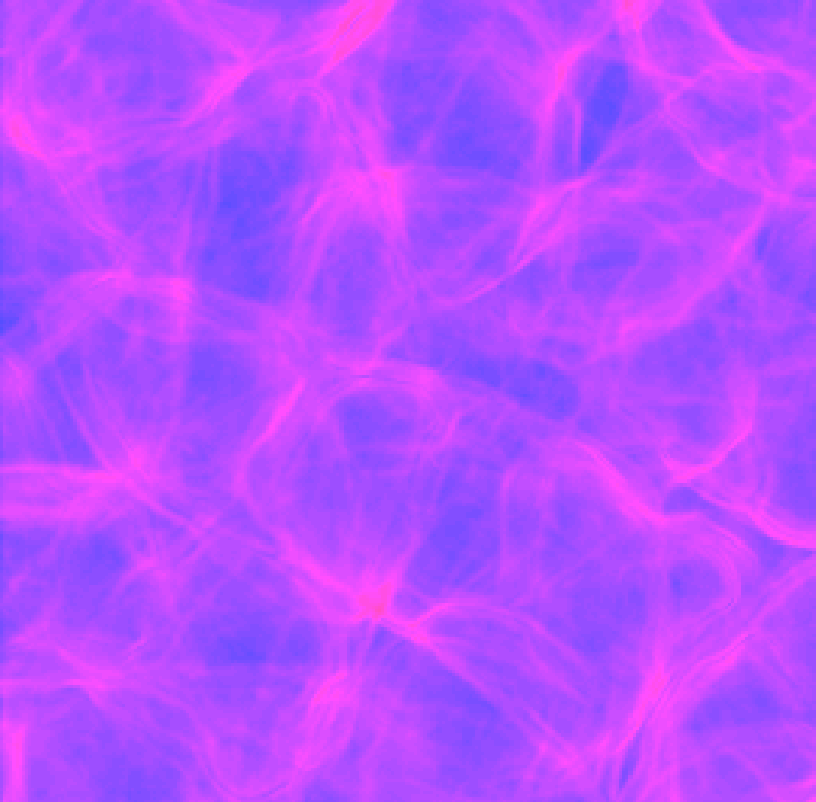}
}
\caption{
Schlieren images in the YZ-plane ($S_x$) of a three-dimensional CJ cellular detonation
at $\rho_0=10^6$~\gcm from runs (a) 1e6C2, (b) 1e6C1, (c) 1e6C,
and (d) 1e6D. Runs 1e6C,C1,C2 were done in the same computational
domain but with different numerical resolutions $n\simeq 12.3$, $24.5$
and $49$, respectively.  Run 1e6D has the same resolution $n\simeq
12.3$ as 1e6C but was done in two-times wider tube. The frame (d) shows the lower left quarter of the 1e6D tube cross-section.  The density gradient increases from blue to red.
\label{Figure-Schlieren-YZ-CC1C2D}
}
\end{figure}

Fig.~\ref{Figure-Schlieren-YZ-CC1C2D} shows a side view Schlieren
image of run 1e6C2 detonation at the same moment of time as in
Figs.~\ref{TP-1e6C2}-\ref{VyVz-1e6C2}.  Both the leading shock and
multiple transverse waves are visible, but due to a large number of
transverse waves, the estimation of the characteristic cell size $l_c$
is impossible. To estimate $l_c$ we use the head-on $S_x$ Schlieren
images.  Figs.~\ref{Figure-Schlieren-YZ-CC1C2D}a--d show $S_x$
images of a fully developed cellular detonation for runs 1e6C2, 1e6C1,
1e6C, and 1e6D, respectively.  The first three simulations are
performed using the same width $W=20.8x_c$ and three different
numerical resolutions, $n_c=12.3$, $24.5$ and $49.0$. The last
simulation is done at the lowest resolution $n_c=12.3$ resolution, but
inside a two times wider computational domain, $W=41.7x_c$.
Fig.~\ref{Figure-Schlieren-YZ-CC1C2D}d shows one-quarter of the
YZ-cross section. 

With increasing resolution, we observe a finer and more detailed
structure of transverse waves.  The figure shows that the cellular
structure is rather chaotic, but irregular detonation cells are
discernible. On average, one can identify $\simeq 3-4$ detonation
cells across the width of the computational domain. The cells seem to
be slightly larger at the lowest resolution but are practically the
same at the two higher resolutions.  From the simulations, we roughly
estimate $l_c\simeq 5 x_c$.  The results obtained at different
numerical resolutions indicate that the numerical resolution does not
significantly affect the estimation of $l_c$. The independence of $l_c$ from
the tube width confirms that the width of the computational domain is
sufficient, and tube geometry does not affect the cell size. The
cell size estimated from the simulations is the intrinsic cell size of
an unconfined freely propagating CJ detonation.

\begin{figure}
\centering
    \subfigure[]{
    \includegraphics[scale=0.360,angle=0]{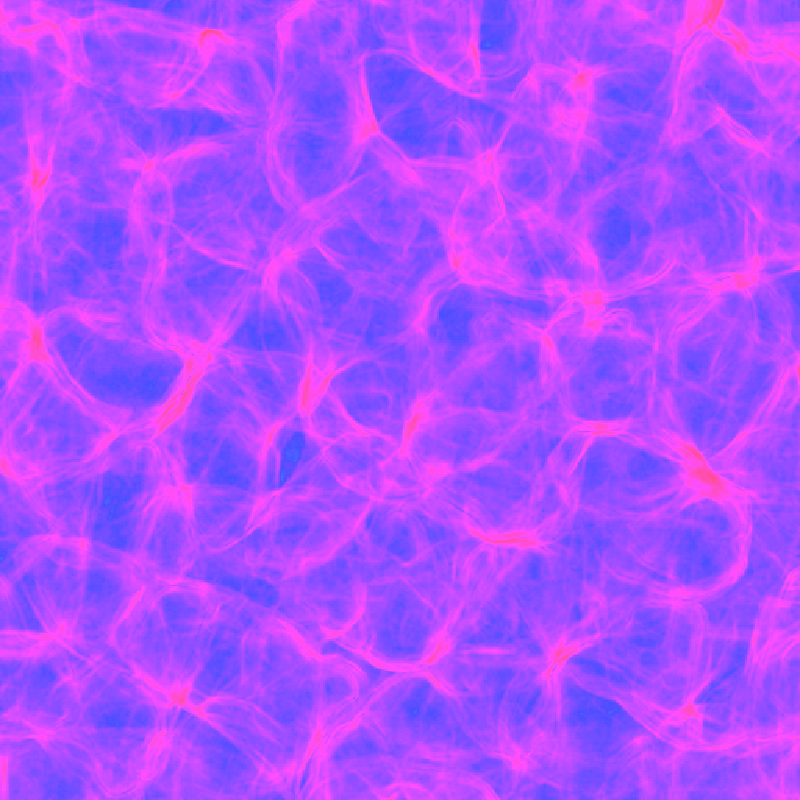}
}
    \subfigure[]{
    \includegraphics[scale=0.380,angle=0]{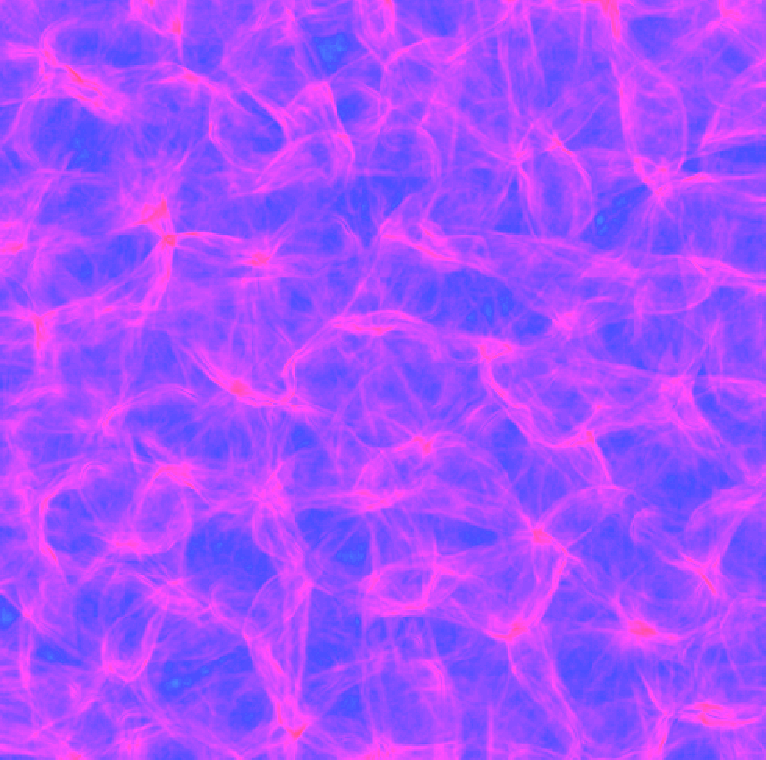}
}
    \subfigure[]{
    \includegraphics[scale=0.362,angle=0]{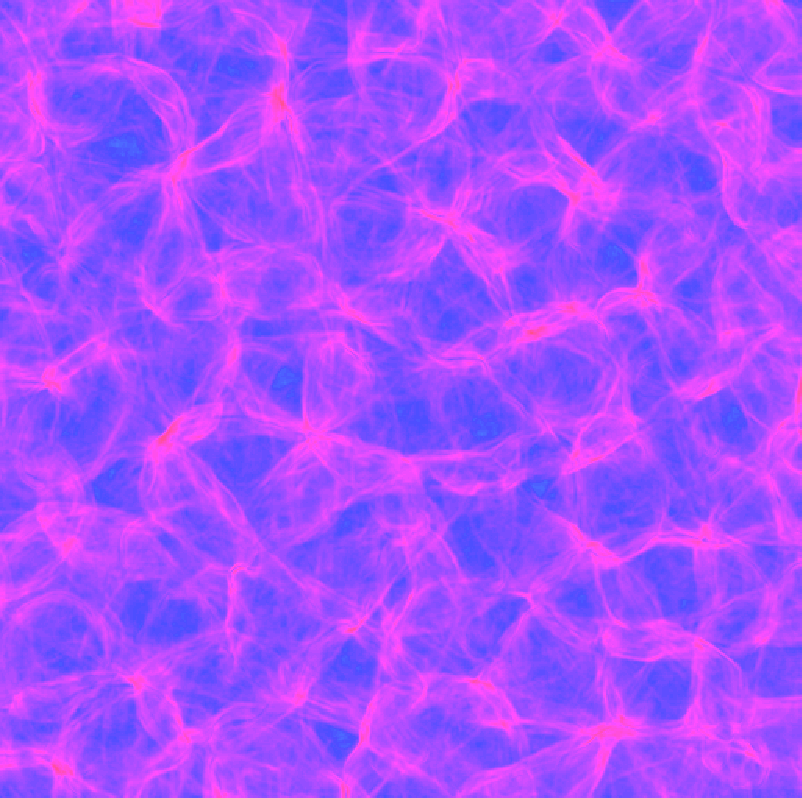}
}
\caption{Schlieren $S_x$ images in the YZ-plane for a three-dimensional  cellular CJ 
detonation at different densities. 
(a) run 1e6D, $\rho_0=10^6$~\gcm, $t=0.046$ s.  
(b) run 7e5B, $\rho_0=7\times 10^5$~\gcm, $t=0.091$ s.  
(c) run 5e6B, $\rho_0=5\times 10^5$~\gcm, $t=0.49$ s.  
Detonation propagates towards the viewer. Frames are scaled to the same relative size.
Values of $t_C$ and $W$ for the runs are given in Table.~\ref{TableSteadyState} and Table.~\ref{TableConstantRuns3D}, respectively.
The density gradient increases from blue to red.
\label{Figure-Schlieren-YZ-5710}
}
\end{figure}

Fig.~\ref{Figure-Schlieren-YZ-5710} compares the cell structure of
detonations propagating at different densities, $\rho_0=5\times 10^5$,
$7\times 10^5$ and $1\times 10^6$~\gcm.  The half-reaction length
of the detonation is a strong function of $\rho_0$ and varies from
highest to lowest density by more than an order of magnitude
(Table~\ref{TableSteadyState}). To compare detonations at different
$\rho_0$ the $S_x$ images were scaled to the same relative size.  The
comparison shows a remarkable similarity between the detonation cells
at various densities. The relation between the detonation cell size
and the half-reaction length of a ZND detonation, is
$l_c\simeq{5}x_c$ and holds for all densities.

\subsection{Two-dimensional detonation}
\label{TwoDimensionalDetonation}

Simulations show a striking difference in the propagation of
two-dimensional detonation waves as compared to a propagation of the
same detonations in three dimensions.  Instead of developing a robust
small-scale cellular structure, two-dimensional detonations are
characterized by quasi-periodic episodes during which the leading
shock weakens and separates from the reaction zone. The layer of
accumulated unburnt fuel is then incinerated by a transverse
detonation wave, which bounces off the sides of the computational
domain.  

\begin{figure}
\centering
\vskip 1cm
\subfigure[]{
     \includegraphics[scale=0.6,angle=0]{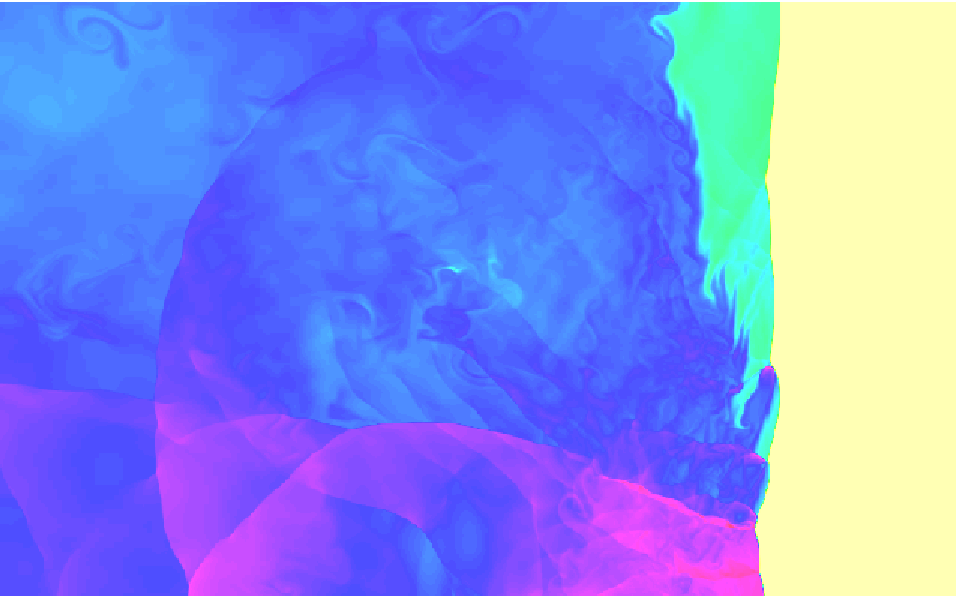}
}
\subfigure[]{
     \includegraphics[scale=0.6,angle=0]{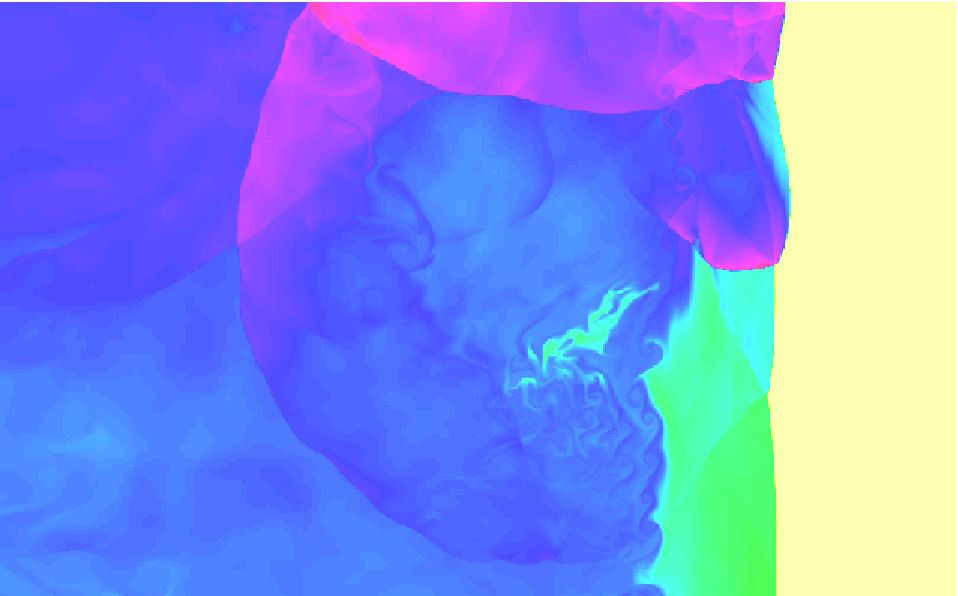}
}
\subfigure[]{
     \includegraphics[scale=0.6,angle=0]{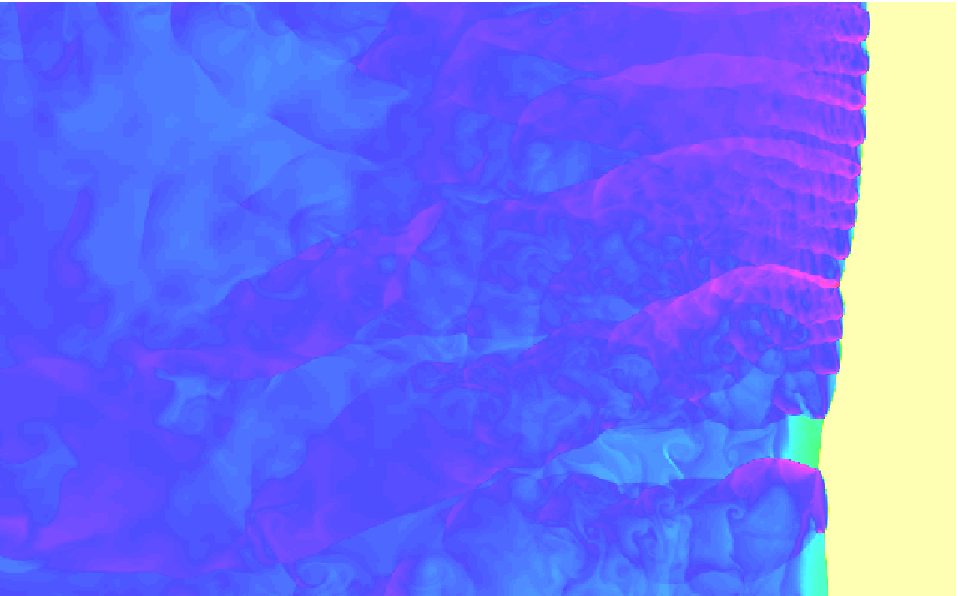}
}
\vskip -6.5 cm
\hskip -7.5cm
\includegraphics[scale=0.3,angle=0]{Scale.eps}
\vskip 2cm
\caption{Two-dimensional cellular detonation at $\rho_0=10^6$~\gcm
(run 1e6C2-2D). The size of the computational domain is L=200 km and W=25 km, as for the three-dimensional run  1e6C2.  Temperature is shown in the XZ-plane passing through
the centerline of the computational domain.  (a) $t=96.2t_C$, 
(b) $t=142.3t_C$ and (c) $t=126.9 t_C$.   The range is $T=\{0-4\times 10^9\}$K.
Color palette is explained in Appendix~\ref{Visualization}.
\label{T-1e6C2-2D}
}
\end{figure}

As an example, Fig.~\ref{T-1e6C2-2D} shows the propagation of a
two-dimensional detonation in a $0.5\ctw+0.5\oos$ mixture at
$\rho_0=10^6$~\gcm.  The
computational setup and numerical resolution 
for this calculation are exactly the same as for the 
three-dimensional run $1e6C2$, except that the third dimension is
eliminated.  In Fig.~\ref{T-1e6C2-2D}a one can see
a layer of unburned fuel accumulating behind the leading shock (light
blue).  A transverse wave just bounced off the lower boundary and
created a section of a highly overdriven detonation propagating to the
right (red).  With time, the overdriven detonation weakens, and the
reaction zone separates from the shock. Fig.~\ref{T-1e6C2-2D}b shows
the newly formed layer of unburned fuel (light blue) being incinerated
in a transverse wave that bounced off the upper boundary. A new
section of an overdriven detonation created by the bounce is visible
near the upper boundary (red). The process repeats itself, roughly, on
a timescale $t_{2D} \simeq 2W/a_s$ where $a_s$ is the sound speed in
shocked matter.  However, the process is not strictly periodic. From
time to time, the entire cross-section may be briefly filled with an
overdriven detonation, but the detonation eventually weakens and
decays. One of these ``overdriven'' episodes is shown in
Fig.~\ref{T-1e6C2-2D}c. Similar behavior was found in calculations
with increased numerical resolution.

The period increases with increasing $W$ until a certain width is
reached where re-initiation of a detonation begins to take place, not
only at the boundaries but also in the middle of the computational
domain, creating an additional transverse detonation wave moving
behind the leading shock.  Re-initiation in the middle of the domain via the Zeldovich gradient mechanism associated with the temperature non-uniformities has long been suggested, and may lead to a re-ignition. However, our  simulations lack the needed resolution to resolve the gradients. 

Note that previous two-dimensional simulations of CO detonation in degenerate matter obtained a clear cellular structure, in most cases for higher background densities than the case shown here, as they focused on the delayed detonation transition density, $\approx  10^7$ \gcm. In particular, \cite{Boisseau1996}  assumed a background initial density of $3\times10^7$ \gcm, \cite{Gamezo1999} explored a range of densities, from $10^6$ \gcm to $3\times10^7$ \gcm, discussing in detail the $5\times10^6$ \gcm case, \cite{Timmes2000} studied the $10^7$ \gcm case, while \cite{Iwata2024} have  included, in their He detonation simulations, a C detonation at $10^6$ \gcm  . In all these studies the initial carbon mass fraction was 0.5, except in \cite{Timmes2000} that assumed 1.0. Only  \cite{Gamezo1999} and \cite{Iwata2024} studied densities as low as our $10^6$ \gcm case.

A basic difference between these two studies and our two-dimensional example, all done at the same initial density, is the energy release. In our  incomplete C-detonations is 0.37 MeV/nuc, which is a factor of $\approx$2 smaller than that of a complete or NSE-detonation, 0.76 MeV/nuc. Thus, $x_C$ is above two and one order of magnitudes greater than those obtained in \cite{Gamezo1999} and \cite{Iwata2024}, respectively.

In our simulation, the spontaneous re-initiation of a detonation may reflect the
intrinsic properties of an unconfined two-dimensional detonation, and
thus the minimal value of $W$ when this happens may represent the
spatial scale of a two-dimensional detonation cell, $l_{c,2D}$.  We
find that $l_{c,2D}$ is significantly larger than the size of a
three-dimensional detonation cell, $l_{c,2D}\simeq 50l_c$. This size  is much greater than that  obtained by other authors, taking  $x_C$ as a reference, we obtain $\approx$250 $x_C$, compared to $\approx$30 $x_C$ and $\approx$10 $x_C$ obtained by \cite{Gamezo1999} and \cite{Iwata2024}, respectively.

\subsection{Effect of the composition}
\begin{figure}
\includegraphics[scale=0.4,angle=0]{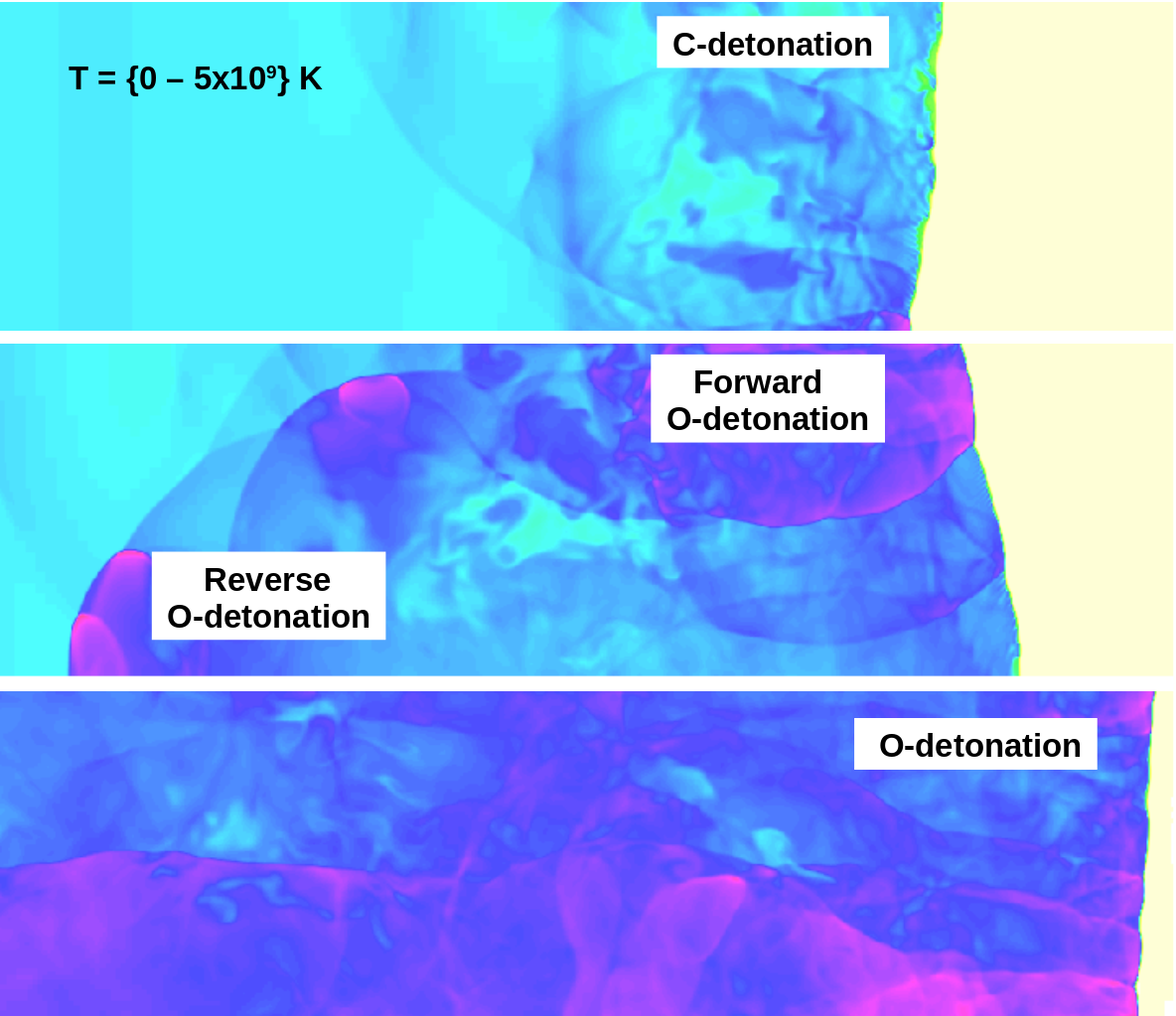}

\vskip -4.5 cm
\hskip 0.2cm
\includegraphics[scale=0.25,angle=0]{Scale.eps}
\vskip1cm
\caption{Temperature evolution of the detonation front for 0.3 C and 0.7 O mass fraction,  at $\rho_0=10^6$~\gcm (similar to run 1e6C2) for  t=50, 123 and 223 $t_C$ in the XZ-plane passing through the centerline of the computational domain. $T=\{0-5\times10^9 K$.
Color palette is explained in Appendix~\ref{Visualization}. 
The rapid transition to explosive O-burning results in high temperatures.}
\label{new}
\end{figure}

\begin{figure}
\includegraphics[scale=0.40,angle=0]{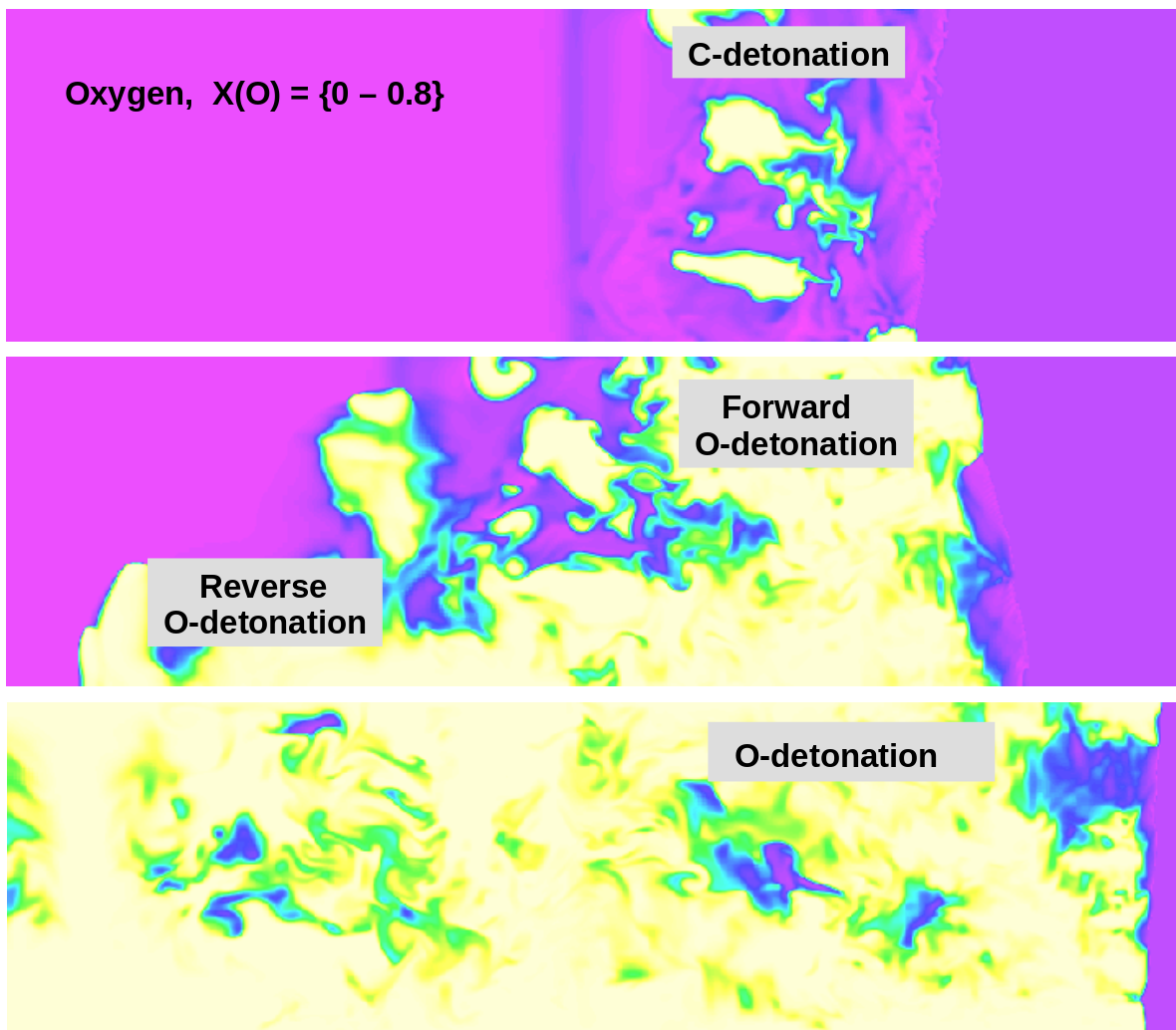}
\vskip -4.5 cm
\hskip 0.2cm
\includegraphics[scale=0.25,angle=0]{Scale.eps}
\vskip1cm
\caption{Mass fraction of \XO, same as Fig. \ref{new}. $\XO=\{0-0.8\}$.
Color palette is explained in Appendix~\ref{Visualization}} 
\label{new2}
\end{figure}

\begin{figure}
\includegraphics[scale=0.40,angle=0]{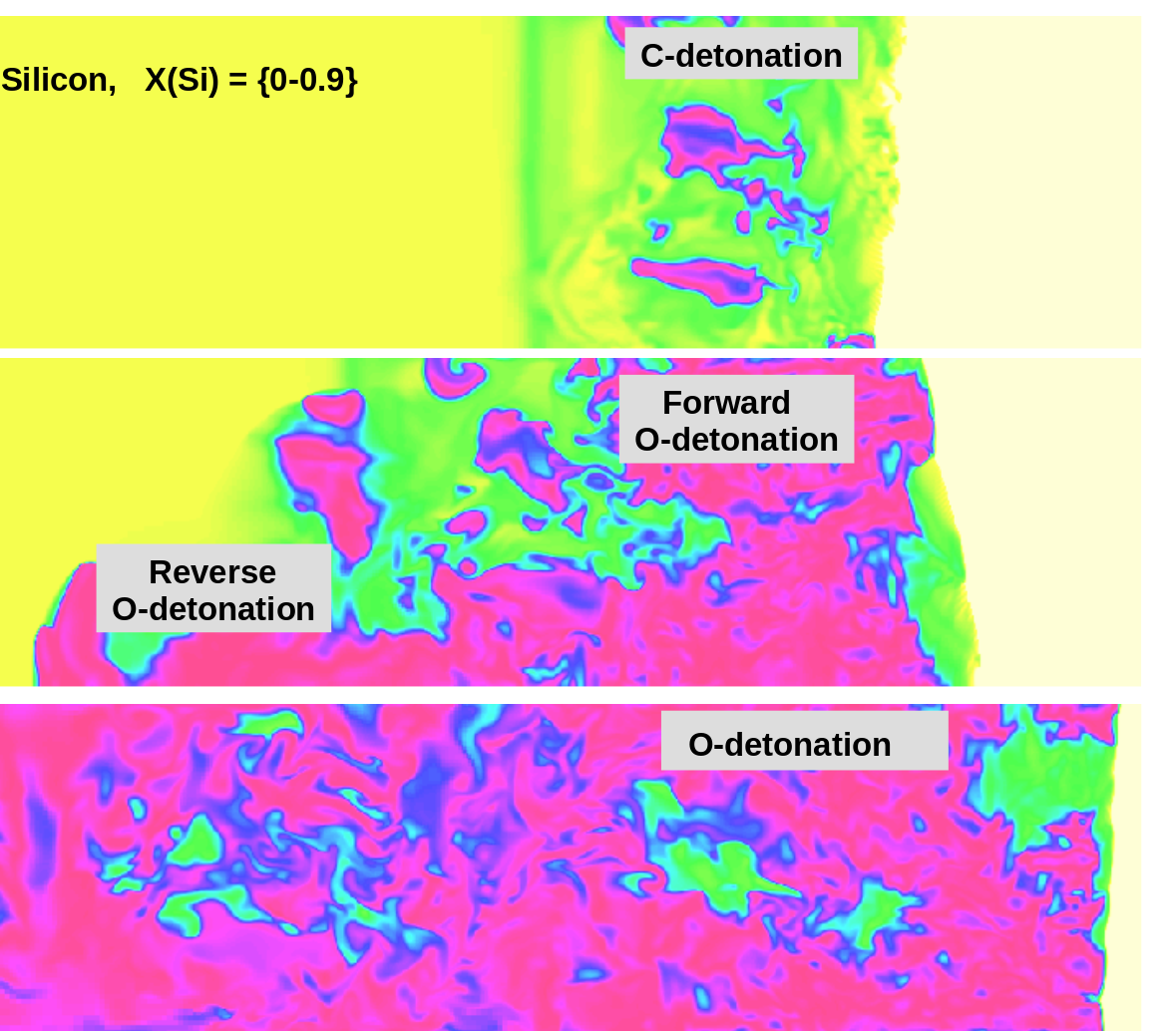}
\vskip -4.5 cm
\hskip 0.2cm
\includegraphics[scale=0.25,angle=0]{Scale.eps}
\vskip1cm
\caption{Mass fraction of \XSi, same as Fig. \ref{new}. $\XSi=\{0-0.9\}$.
Color palette is explained in Appendix~\ref{Visualization}} 
\label{new3}
\end{figure}

3-D simulations show the sensitivity of the detonation burning on the 
abundances or, more precisely, the C/O ratio in the regime of distributed burning. In Figs. \ref{new}, \ref{new2} and \ref{new3}, the evolution
of the detonation is shown at t=92, 143, and 226 $t_c$ for the same parameters of model 1e6C2 with $\rho_c=10^6$~\gcm, but a mixture of 0.3 C and 0.7 O.
The result is qualitatively different from the case of C/O=1 \citep{kd15}. The average post-shock temperature in detonations with \XC = 0.3 is lower
compared to a detonation with \XC = 0.5 due to a lower energy release, 0.22 and 0.37 MeV/nucleon, respectively, in the leading C12 + C12 reaction. As a result the detonation zone is more unstable and is characterized by a higher amplitude of temperature fluctuations. The peak temperature near the triple-points becomes so high that the shock
collisions trigger O-burning which converts matter to Si-group elements and almost doubles the energy
release.  A forward 
and backward O-detonation develops, so the O-burning quickly spreads through the reaction zone. 
The results are few inhomogeneities remaining of C and O and a reduction of layers produced
by incomplete carbon burning.  

Thus, in C-rich mixtures, $\XC\geq0.5$, the detonation propagates as a quasi-steady detonation but in carbon poor mixtures,  $\XC\leq0.3$, the detonation must be treated in a fully non-stationary manner \citep{kd15}.
 
 In close to $M_{Ch}$ mass explosions, assuming  accretion of H/He rich matter from a companion, and that the whole WD is completely mixed before the explosion, the average \XC is expected to range from above 0.4 to 0.5 (C/O from above 0.6 to 1), depending on initial mass and metallicity. Note that \XC could be much larger if the outer layers are not mixed. 

While, for secular and dynamical mergers, lower \XC may be expected, from above 0.3 to 0.4 (C/O from above 0.4 to 0.7), depending on initial WD masses and metallicities. This is due just to the fact that two WDs, instead of one, contribute with the C-poor material coming from the central convective He-burning stellar phase. Central He-burning is a long phase, in which there is enough time for alpha captures on \ctw, thus, decreasing \ctw and producing \oos \citep{Dominguez_etal_2001, Dominguez2003, Bravo_etal_2010, Khokhlov2012, Piersanti2019}.


                                                                         \section{Discussion and Conclusions}
\label{Conclusions}


This paper presents results of  three-dimensional numerical simulations of CO-detonations in a tube with physical and chemical properties similar to the outer layers of SNe~Ia. We study the impact of dimensionality comparing  with two- and three-dimensional numerical simulations. We also analyze the  results assuming different background densities and chemical mixtures. 

\begin{itemize}

\item Three-dimensional detonations appear very robust in all the simulations. The detonation cells are very similar at the different studied densities, 0.5 to 1 $\times 10^6$~\gcm. All show detonation cell sizes above 5 times the carbon  half-reaction length, $x_C$ (which is defined as the distance from the leading shock to a point where half of initial \ctw is consumed).

\item The average three-dimensional structure resembles the steady-state one-dimensional reaction zone. The most remarkable differences are that \ctw is burnt slightly more rapidly and slightly more \siteg is produced in the three-dimensional models.

\item In the three-dimensional simulations, non-uniformities of products of C-burning, like Si, Ne or Mg,  are present at up to 20-30 $x_C$ of the leading shock, while C non-uniformities rapidly burn out and remain closer  to the leading shock, within  5 $x_C$, with no C at further distances.

\item Multidimensional  simulations show that properties of the same detonation are strikingly different when it propagates in three-dimensions and  in two-dimensions. We find that in two-dimensional simulations, instead of developing a robust small-scale cellular structure, the leading shock weakens and separates periodically from the
reaction zone. This behavior is caused
by the intrinsic instability of a C-detonation to longitudinal perturbations. The estimated  detonation cell sizes  are above 50 times larger than in three-dimensional simulations. Two-dimensional simulations may over-predict the abundances of intermediate mass elements and of residual \ctw. Thus, two-dimensional models of Type Ia supernovae involving detonation waves need to be re-investigated using three-dimensional simulations. 

\item This dependence of the problem on the dimensionality suggests that the outcome of incomplete detonations in SNe~Ia should depend on the geometry of the exploding star which may depend on the progenitors, on the direction of the propagation with respect to the density gradients, and/or whether it is deformed by rotation.

\item We find incomplete C-detonations to be highly unstable. To treat a detonation as a quasi-steady wave, two conditions must be fulfilled: (a) large separation of the half-reaction spatial scales of C, O and Si-burning, and (b) sufficient stability to avoid the transition to an O-detonation. The first condition is usually met in C-detonations when the C abundance mass fraction $\geq$0.5, while we find the second to be violated in carbon poor mixtures. 

\item If carbon abundance $\XC\leq0.3$ a rapid transition to O-detonation occurs. This is due to the increasing instability of the incomplete carbon detonation due to the lower carbon abundance and lower energy release. The amplitude of the temperature fluctuations increases exponentially with the $\beta$ parameter  \citep[see][]{dominguez2011} for the stability study, being $\beta$=16.6 in our 0.5C + 0.5O model, and 20 for the 0.3C + 0.7O model. Thus, chemical composition of the outer layers plays a crucial role in the explosions. In SNe~Ia,  initial low carbon abundance closed to the surface is expected when two or three WDs are involved as progenitors. Note that  the final masses  of the exploding WDs  could be  sub-Chandrasekhar or  nearly Chandrasekhar. In these cases  little or no oxygen will remain in the outer layers after the explosion. While an initial rich carbon outer zone, expected when H/He is accreted and converted to CO, would  finally show larger oxygen abundances.

\item 
Chemical inhomogeneities are rapidly burned away in the region of explosive oxygen burning because the burning time scales are shorter than the expansion time scales. However, the burning time scales become larger in layers of incomplete carbon burning, and will `freeze' out in the outer layers.

\item Mapping the computational domain of the simulations  into the layers with incomplete carbon-burning of spherical explosion models \citep[][their Fig. 3]{Hoeflich_etal_2019}, final expansion velocities of 
$\approx 20,000 -30,000$ \kmps are exposed in the ejecta during the first few days after the explosion. The scales of the inhomogeneities correspond to about 1000 \kmps. 

\item 
Cellular instabilities can be expected in the entire C-layers and affect the flame propagation. However, the chemical inhomogeneities are burned away at layers with higher densities because the hydrodynamical time scales for the explosion are larger than the burning time scales $t_c$.  

\item
The results of inhomogeneities is a `picket-fence' structure consisting of burned and unburned layers. The result is an increase in the luminosity when the photosphere passes the corresponding layers \citep{2023ApJ...948...10A}. 
Early bumps in the LCs and polarization spectra coincide with the low-density domain and may be caused by chemical inhomogeneities, as discussed in Sect.~\ref{Preamble}.



\item Recently, the mechanism of turbulent deflagration-to-detonation transition has been established \citep{2019Sci...366.7365P}. However,
the current study may also be relevant to 
 detonation initiation, including multi-spot DDTs, since critical conditions for
initiation and for the DDT correlate with basic properties of the detonation waves. 

\end{itemize}  

In summary, the modeling of low density incomplete detonation in SNe~Ia, especially in low carbon matter as may result when two or three WDs are involved as progenitors, requires time-dependent, three-dimensional  reactive flow simulations with numerically-resolved C and O-burning zones.

This work establishes detonation  instabilities and cell structure as important ingredients 
in the physics of thermonuclear supernovae. Now, models, including density gradients and full radiation-hydrodynamics or radiation-magnetic-hydrodynamics simulations \citep{Remming_Khokhlov_2014,Hristov_etal_2018}, are needed. 

\noindent
{\bf Acknowledgements:}
The co-authors are grateful to Alexei for many years of collaboration and friendship. 
The work was carried out within the NSF project “Collaborative
research: Three-Dimensional Simulations of Type Ia Supernovae:
Constraining Models with Observations”  supported by the
NSF grant AST-0709181; I.D acknowledges funding from the project PID2021-123110NB-I00 financed by the Spanish MCIN/AEI /10.13039/501100011033/ \& FEDER A way to make Europe, UE, and the Spanish Ministry for Education Mobility Programme within the framework of the National Plan I+D+I
2008-2013. We thank the staff of the ALCF at the Argonne National Laboratory, especially Mr. C. Bacon (ALCF), for their help with the code development.

\bibliographystyle{aasjournal} 

\appendix

\section{Visualization}
\label{Visualization}

Two-dimensional fields of physical variables are presented using a
standard color palette. represented by a color bar in each
figure. Color bars in each figure are marked with a scale variable,
$c$, which ranges from zero to one, $0\leq c\leq 1$.  A displayed
physical variable, say, $f$, is related to $c$ as
\labeq{ColorCoding}{
           f = f_{min} + c \cdot ( f_{max} - f_{min} ), \quad 
c = \frac{f - f_{min}}{f_{max} - f_{min}}, 
}
where $f_{min}$ and $f_{max}$ are the minimum and maximum values of
$f$. They are listed in figure captions.  For example, in
Fig.~\ref{TP-1e6C2}a the range of temperature is $T=\left\{0 -
4\times10^9\right\}$~K.  The correspondence between $T$ and $c$ is
then $T=4\times10^9\cdot{c}$. In Fig.~\ref{TP-1e6C2}b the
correspondence between pressure $P$ and $c$ is
$\lg(P)=22.0+2.0\cdot{c}$.

\begin{figure}
\centering
\includegraphics[scale=0.6,angle=0]{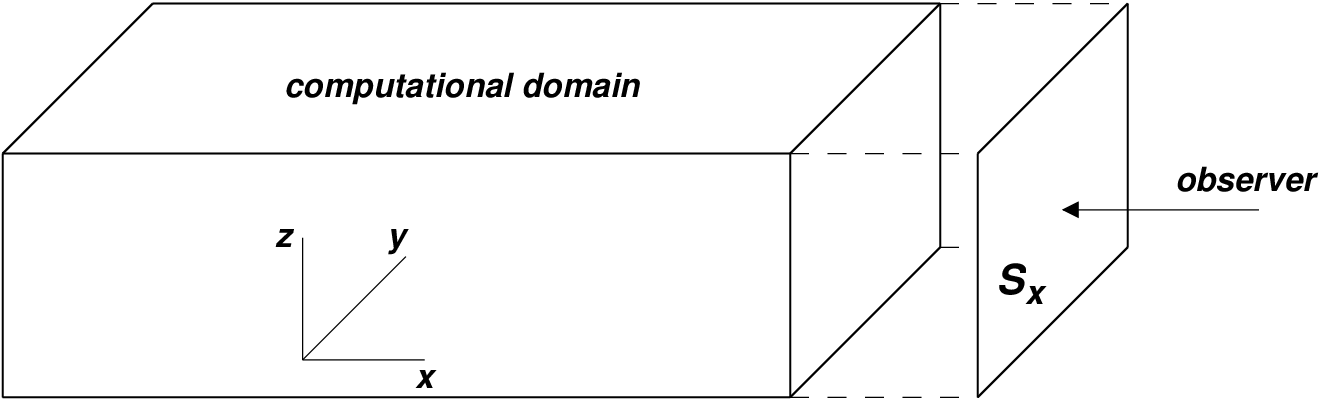}
\caption{Schematic of Schlieren visualization. Detonation propagates
through the computational domain from left to right. A frontal (head-on) Schlieren image $S_x$ is generated by integrating along the X-axis of an absolute value of the orthogonal component of the density gradient, Equation \eqref{SchlierenFrontView}.
\label{Figure-SchlierenViz}
}
\end{figure}

Numerical Schlieren images are used to visualize three-dimensional fields with large density gradients, such as shocks, reaction fronts, and contact discontinuities.  The images are generated by an absolute value of a component of a density gradient orthogonal to a viewing direction along the line of sight (Fig.~\ref{Figure-SchlierenViz}).
\labeq{SchlierenFrontView}{ 
S_x(y,z) = \int_0^L \sqrt{
\left(\pd{\rho}{y}\right)^2 + \left(\pd{\rho}{z}\right)^2 } dx. 
} 
Head-on images do not detect the leading shock and provide an unobscured view of shocks and reaction fronts moving in transverse directions.

\section{Stiff integration of nuclear kinetics}
\label{YASS}

The code uses a cell-based AMR based on a parallel fully threaded tree structure adaptive mesh refinement (AMR) reactive flow fluid
dynamic code ALLA~\citep{Khokhlov1998}.  
Reaction terms were coupled to
fluid dynamics via operator-splitting.
During the hydrodynamical
sub-step the Euler equations were integrated with $\V{\cal R} = \dot q
= 0$ using an explicit, directional-split, second-order accurate,
Godunov-type conservative scheme with a Riemann solver.
The time step $\Delta t = CFL \cdot \frac{ \Delta x}{a_s}$, where and $a_s$ is the velocity of sound and $\Delta x$ is the 
minimum cell size selected using a Courant number $CFL=0.7$.  
During the chemical sub-step, the kinetic equations
together with the equation of energy conservation,
\labeq{DVDY}{
  \D{\V{\cal Y}}{t} = \V{\cal R}, \quad \D{\epsilon}{t} = \dot q, 
               \quad\rho=const.,
}
were integrated using a stiff solver, YASS (Yet Another Stiff Solver, developed by A. Khokhlov, \cite{KhokhlovNRL1999}), with adjustable sub-cycling when necessary to
keep the relative accuracy of integration $10^{-3}$ for individual species with ${\cal{Y}}_i > 10^{-5}$.  The solver conserved the total baryon number and the sum of internal and nuclear binding energy exactly. Below, we briefly summarize its general properties. 

General requirements for a numerical method for integration \eqref{Euler-rho} - \eqref{Euler-rhoY} 
easily follow from the analysis of temporal and spatial scales of the carbon-oxygen detonation. 
The temporal resolution of multi-species reactive flow simulations must be sufficient for accurately resolving critical reaction-kinetic time scales associated with the nuclear energy release and
production of major nuclear species. The spatial resolution of the simulations must match 
their temporal resolution to account for the hydrodynamic interaction of
the nuclear reactions and the fluid motions.
Including networks requires a much higher temporal and spatial numerical resolution and computational resources as compared to simulation with simplified progress-variable kinetics and/or flame capturing and 
level-set descriptions of flame propagation.
 The increased temporal resolution is needed to account for a wide range of reaction time scales. The correspondingly increased spatial resolution is necessary to represent the interplay of nuclear kinetics and fluid flow properly. Typically, nuclear reaction rates vary rapidly with time and in space. For example, the $\ctw+\ctw$ reaction rate is usually the highest immediately behind the detonation shock and becomes much smaller away from the shock where \ctw has been exhausted.
Therefore, the resolution requirements vary rapidly with time and space as well. This calls for 
an adaptive mesh refinement (AMR) approach to
the simulations. AMR is able to concentrate computational resources where they are needed and to accelerate the computations significantly. Additional AMR savings come from increasing
 the numerical resolution near sharp features of a flow, such as shock waves, discontinuities, gradients of physical variables, and regions of high vorticity.

In this paper, we use a new massively parallel AMR code HSCD for first-principles reactive Navier-Stokes numerical simulations of high-speed combustion and 
detonation phenomena in terrestrial gases. 
The code is based on the Eulerian reactive flow AMR code ALLA, which has been used for reactive flow terrestrial and astrophysical combustion phenomena in the past \citep{Khokhlov1998}. 

Eulerian high-speed combustion and a massively parallel version with an adaptive mesh refinement 
immediately after each timestep are required because the simulation reaction timescales vary with spatial resolution. If $\Delta_{t,min}$ $\varepsilon$ $\tau_{n,min}$ is the required temporal resolution of numerical integration associated with the shortest relevant reaction timescale  ($\tau_{n,min}$, of reactant n), 
then the corresponding minimal spatial scale should be of the order $\Delta_{x,h} \simeq a_s \Delta_{t,min}$ $\varepsilon$  $\tau_{n,min}$, and $\varepsilon < 1$ is based on the consideration of the accuracy of kinetic numerical integration.  
For an explicit hydrodynamical code, the hydrodynamical time step $\Delta_t$,
must be limited by the CFL stability condition, 
\labeq{CFL2}{
              \Delta_t < \Delta_{t,CFL} = \min \left( \frac{\Delta_{x,h}}{\max (a_s + \vert u\vert)}\right).
} 
where u is the fluid velocity relative to the shock, and the maximum taken over all three directions. 

Hydrodynamical simulations of SNe~Ia with multi-species reaction networks, such as the $\alpha$-network used in this paper or more extensive networks including
more species require much more computing power than simulations with simplified model networks.
There are several reasons for this. One is the cost of computing nuclear kinetics.
The right-hand sides of \eqref{Euler-E}, \eqref{Euler-rhoY} in this case, become much more expensive compared to those of quasi-networks. The latter may contain $\simeq 1-4$ pseudo-kinetic variables and have simple right-hand sides. Secondly, compared to the quasi-networks, the multi-species networks are 
extremely stiff. Stiffness of an $\alpha$-network, for example, reaches $\simeq 10^6 - 10^{13}$  in conditions of SNe~Ia. 
As a result, the numerical integration requires unconditionally stable implicit methods and must involve evaluation of the Jacobian of the right-hand sides of the kinetic equations. This further raises the cost of kinetic integration. Such astrophysical networks are impossible to integrate using asymptotic methods (see e.g. \cite{Oran2000}.

\clearpage

\begin{deluxetable}{ccccccccccccccc}
\tablecolumns{9} 
\tablewidth{0pc} 
\tablecaption{Properties of Chapman-Jouguet C-detonations\label{TableSteadyState}}
\tablehead{ 
	      \colhead{$X_C$~$^{(a)}$}
            & \colhead{$\rho_0$~$^{(b)}$}
            & \colhead{$\frac{\rho_s}{\rho_0}$~$^{(c)}$}  
            & \colhead{$T_s$~$^{(d)}$}   
            & \colhead{$D_{CJ}$~$^{(e)}$}   
            & \colhead{$x_C$~$^{(f)}$}   
            & \colhead{$t_C$~$^{(g)}$}   
}
\startdata
$0.5$ &0.50 & 4.57 & 2.06 & 0.85 & $1.9\times 10^6$ & $8.6\times 10^{-3}$  \\ 
$0.5$ &0.70 & 4.77 & 2.18 & 0.92 & $3.6\times 10^5$ & $1.6\times 10^{-3}$  \\ 
$0.5$ &1.00 & 4.35 & 2.27 & 0.87 & $1.2\times 10^5$ & $5.2\times 10^{-4}$  \\
%
%
\hline
$0.3$   &1.00 & 3.86 & 1.74 & 0.72 & $2.7\times 10^7$ & $1.3\times 10^{-1}$  \\
$0.3$   &3.00 & 3.47 & 1.92 & 0.75 & $9.8\times 10^5$ & $4.2\times 10^{-3}$  \\
$0.3$   &10.0 & 2.99 & 2.07 & 0.80 & $5.0\times 10^4$ & $1.8\times 10^{-4}$  \\
\enddata
\tablenotetext{(a)}{~~~$ X_C$ - initial mass fraction of \ctw. 
Energy release is $q_0=0.37$ MeV/nucleon and $q_0=0.22$ MeV/nucleon
for $X_C=0.5$ and $X_C=0.3$ compositions, respectively.}

\tablenotetext{(b)}{~~~$\rho_0$ - background density in $10^6$~\gcm.}

\tablenotetext{(c)}{~~~$\rho_s$ - post-shock density.}

\tablenotetext{(d)}{~~~$T_s$ - post-shock temperature in $10^9$~K.}

\tablenotetext{(e)}{~~~$D_{CJ}$ - Chapman-Jouguet detonation velocity in $10^9$ \cmps.}

\tablenotetext{(f)}{~~~$x_C$ - half-reaction length of a detonation in cm.}

\tablenotetext{(g)}{~~~$t_C$ - half-reaction time of a detonation in sec.}

\end{deluxetable}


\clearpage

\begin{deluxetable}{lcccccccccccccc}
\tablecolumns{11} 
\tablewidth{0pc} 
\tablecaption{Three-dimensional simulations in $0.5\,\ctw+0.5\,\oos$ mixture\label{TableConstantRuns3D}}
\tablehead{  \colhead{Run}
            & \colhead{$\rho_0$~$^{(a)}$}
            & \colhead{$L$~$^{(b)}$}  
            & \colhead{$\left(\frac{L}{W}\right)$~$^{(c)}$}  
            & \colhead{$l_{max}$~$^{(d)}$}  
            & \colhead{$n_{1/2}$~$^{(e)}$}  
            & \colhead{$\left(\frac{W}{x_C}\right)$~$^{(f)}$}   
            & \colhead{$\left(\frac{y_p}{W}\right)$}
            & \colhead{$\left(\frac{z_p}{W}\right)$~$^{(g)}$}
}
\startdata
5e5B   & 0.5        &  32.0      & 4 & 11 & 12.3 & 20.8  & 0.25   & 0.625  \\
7e5B   & 0.7        &  6.0       & 4 & 11 & 12.3 & 41.7  & 0.25   & 0.625  \\
1e6A1  & 1.0        &  2.0       & 8 & 12 & 24.5 & 20.8  & 0.25   & 0.5    \\
1e6B   & 1.0        &  2.0       & 4 & 11 & 12.3 & 41.7  & 0.25   & 0.5    \\
1e6C   & 1.0        &  2.0       & 8 & 11 & 12.3 & 20.8  & 0.18   & 0.444  \\     
1e6C1  & 1.0        &  2.0       & 8 & 12 & 24.5 & 20.8  & 0.18   & 0.444  \\
1e6C2  & 1.0        &  2.0       & 8 & 13 & 49.0 & 20.8  & 0.18   & 0.444  \\
1e6D   & 1.0        &  2.0       & 4 & 11 & 12.3 & 41.7  & 0.18   & 0.444  \\
\enddata
\tablenotetext{(a) }{
~~$\rho_0$ - background density in $10^6$~\gcm.}
\tablenotetext{(b) }{
~~$L$ - lenght of the computational domain in $10^7$ cm.}
\tablenotetext{(c) }{
~~$W$ - width of the computational domain.}
\tablenotetext{(d) }{
~~$l_{max}$ - maximum level of mesh refinement.} 
%
\tablenotetext{(e) }{~~$n_{1/2}$ - number of computational cells per a
half-reaction length $x_C$.
}
\tablenotetext{(f) }{~~$W/x_C$ - 
number of half-reaction lengths $x_C$ per width W.
}
\tablenotetext{(g) }{~~$y_p$ and $z_p$ are Y- and Z-coordinates of a spherical perturbation,
respectively. Radius, $r_p=\frac{1}{4}x_C$,  and X-coordinate,
 $x_p=0.9L-\frac{2}{3}x_C$, of a spherical perturbation are the same in all runs.}
\end{deluxetable}

%

\clearpage


\clearpage


\clearpage

\clearpage


\clearpage
\clearpage


\clearpage


\clearpage


\clearpage


\clearpage


\clearpage
\vskip 1cm

\clearpage


\clearpage

\clearpage


\clearpage


\clearpage
\clearpage

\clearpage

\clearpage
\clearpage

\end{document}